\documentclass[journal]{IEEEtran}
\usepackage{cite}
\newcommand{\qvec}[1]{\textbf{\textit{#1}}}
\usepackage{cite}

\ifCLASSINFOpdf
 \usepackage[pdftex]{graphicx}
\else
\usepackage[dvips]{graphicx}
\fi

\usepackage{amsmath,amssymb,amsfonts,bm,setspace}
\usepackage[ruled,lined,commentsnumbered]{algorithm2e}
\usepackage{bbm}

\usepackage{tikz}
\usetikzlibrary{shapes.geometric, arrows}
\usepackage{pgfplots}
\usepackage{relsize}

\usepackage{tabularx,ragged2e,booktabs,caption,array}
\usepackage{lipsum}

\usepackage{tikz}
\usetikzlibrary{shapes.geometric, arrows}
\usepackage{pgfplots}
\usepackage{relsize}
\usepackage{xcolor}

\usepackage{todonotes}
\usepackage{arydshln}
  \usepackage[caption=false,font=footnotesize]{subfig}
\usepackage{url}

\newcolumntype{b}{X}
\newcolumntype{s}{>{\hsize=.5\hsize}X}
\usepackage{tabularx}
\usepackage{multirow}
\begin{document}

\input{tikZstyle.tkZ}
\title{Sparsity-aware Robust Community Detection (SPARCODE)}

\author{ 
Aylin~Ta{\c{s}}tan,~\IEEEmembership{Student Member,~IEEE,}
        Michael~Muma,~\IEEEmembership{Member,~IEEE,}\\
        and~Abdelhak M.~Zoubir,~\IEEEmembership{Fellow,~IEEE}
\thanks{The authors are with the Signal Processing Group, Technische Universität
Darmstadt, Darmstadt, Germany (e-mail: atastan@spg.tu-darmstadt.de; muma@spg.tu-darmstadt.de; zoubir@spg.tu-darmstadt.de).}
}


\maketitle

\begin{abstract}
Community detection refers to finding densely connected groups of nodes 
in graphs. In important applications, such as cluster analysis and 
network modelling, the graph is sparse but outliers and heavy-tailed 
noise may obscure its structure. We propose a new method for 
Sparsity-aware Robust Community Detection (SPARCODE). Starting from a 
densely connected and outlier-corrupted graph, we first extract a 
preliminary sparsity-improved graph model where we optimize the level of 
sparsity by mapping the coordinates from different clusters such that 
the distance of their embedding is maximal. Then, undesired edges are 
removed and the graph is constructed robustly by detecting the outliers 
using the connectivity of nodes in the improved graph model. Finally, 
fast spectral partitioning is performed on the resulting robust sparse 
graph model. The number of communities is estimated using modularity 
optimization on the partitioning results. We compare the performance to 
popular graph and cluster-based community detection approaches on a 
variety of benchmark network and cluster analysis data sets. 
Comprehensive experiments demonstrate that our method consistently finds 
the correct number of communities and outperforms existing methods in 
terms of detection performance, robustness and modularity score while 
requiring a reasonable computation time.
\end{abstract}
\vspace{1mm}

\IEEEpeerreviewmaketitle

\vspace{6mm}

\section{Introduction}
\vspace{2mm}

\IEEEPARstart{I}{nferring} a graph model from empirical observations is a fundamental data science task, and a large number of graph construction algorithms have been proposed, e.g. \citeform[1]-\citeform[4]. In network modelling, graphs are used to represent the interactions between components of a system \cite{networkmodelling}, and in cluster analysis, the similarity between features can be expressed by a weighted graph \cite{sparserepimage}, \cite{RadarConf}. Graph models play a crucial role, for example, in subspace learning \cite{SparseSubspace},\cite{subspacelearning}, manifold learning \cite{manifold1}, \cite{manifold2}  and semi-supervised learning \cite{semisupervised}. In particular, the inference of  a graph model forms the basis of graph partitioning \cite{graphpartition1}, \cite{graphpartition2} and community detection algorithms, which has been a very active area of research in recent years \citeform[13]-\citeform[19].
\vspace{1mm}

Community detection refers to finding densely connected groups of nodes, which helps to deduce the underlying structure and relationships that are inherent to the data. A rather important and typical situation is when the data is corrupted by heavy-tailed noise and outliers \citeform[20]-\citeform[31]. 
This may lead to a performance degradation for popular graph-based community detection methods, e.g. \citeform[13]-\citeform[15],
\cite{Louvain} using modularity optimization, which is the most widely used objective function for partitioning \cite{Newmanmod}. A reason for the performance loss is that these methods apply optimization directly on a graph that is possibly corrupted with undesired edges that may be caused by outliers and noise. Therefore, in real-world settings with large and densely connected graphs, including graphs with a considerable amount of outliers, classical community detection methods based on modularity optimization are not be capable of recovering a graph that well-represents the underlying structure of the clean data. For example, methods that use pair-wise Euclidean distances, such as $k$-nearest neighbor and $\epsilon$-ball, are particularly sensitive to noise and outliers \cite{l1graph}. 
\vspace{1mm}

Graph-construction algorithms that use sparse representation may provide a performance gain compared to Euclidean distance-based methods \cite{l1graph}, \cite{elasticnethypergraph}, \cite{sparserepimage}, \cite{sparsegraphest} and sparsity-based characterization of locality relations can be valuable for cluster analysis \cite{RadarConf}, \cite{manifold2}, \cite{sparsityinclustering}. However, the performance of these clustering methods is sensitive to the level of sparsity which is essential in graphs. Sparsity in graphs has been extensively researched, e.g. in terms of the geometry of graphs and there are several approaches \cite{LeightonRao}, \cite{Geometryflows} for sparsity approximation. Nevertheless, sparse methods are also affected by outliers, and determining the suitable level of sparsity becomes especially challenging. 
Furthermore, finding such a sparse embedding is often NP hard \cite{AppSparsestcut2}.
\vspace{1mm}

A popular approach to promote sparsity is the LASSO regularization, which brings a relaxation for increasing sparsity on a graph without necessitating dimension reduction \cite{Glasso}. However, the performance of the graphical LASSO critically depends on the selection of the penalty parameter that controls the sparsity of the graph.  It is well known that the selection of the penalty parameter is a challenging problem in both semi-supervised \cite{graphbasedfeaselect}, and unsupervised settings \cite{penaltyselect1}, \cite{penaltyselect2}; and often supervised approaches \cite{supervisedpenalty1}, \cite{Glasso} or  neighborhood selection \cite{neighborselect}, are used. An interesting approach to sparsity control using the penalty parameter for graphical LASSO was made in \cite{sparsitypenalty} by utilizing knowledge of the number of connected components of the graph. The approach controls the sparsity based on the a priori knowledge of the number of clusters, which may be difficult to estimate in the presence of outliers. To the best of our knowledge, robust sparsity control for graphical LASSO has not been applied to community detection.

To address the above challenges, we propose a new method for Sparsity-aware Robust Community Detection (SPARCODE). The method begins with a densely connected graph and produces a preliminary sparsity improved graph, obtained via an $\ell_1$-penalized precision matrix estimation. We propose a method to optimize the penalty to provide a mapping of the coordinates from different clusters in such a way that they are embedded as far as possible on the real line. Then, undesired and negligible edges are removed from the sparsity improved graph model and the graph construction is performed in a robust manner by detecting the outliers based on connectivity of nodes in the improved sparse graph model. Finally, fast spectral partitioning is performed on the outlier-free vertices of the robust sparse graph model. The number of communities is estimated using modularity optimization on partitions.


The paper is organized as follows. Section \ref{sec:Preliminaries} comprises the basic concepts and problem formulation. The proposed community detection algorithm is detailed in Section \ref{sec:Proposed}. Section \ref{sec:Experimental} demonstrates the community detection performance of SPARCODE in comparison to graph-based and cluster-based community detection
algorithms using real-world and simulated data. Finally, conclusions are drawn in Section \ref{sec:Conclusion}.
\section{Preliminaries and Problem Statement}\label{sec:Preliminaries}
\subsection{Summary of Notations}
Lower and upper-case bold letters denote vectors and matrices, respectively; $\mathbb{R}$ denotes the set of real numbers; $\mathbb{Z}^{+}$ denotes the set of positive integers; $\mathbb{S}^{o}_{++}$ denotes the set of symmetric nonnegative definite matrices; $|x|$ denotes the absolute value of $x$; $\|\qvec{x}\|$ denotes the norm of vector $\qvec{x}$, e.g. $\|\qvec{x}\|_1$ is $\ell_1$ norm; $\qvec{x}^{(s)}$ denotes sorted vector $\qvec{x}$ where the sorting is in ascending order; $\mathrm{med}(\qvec{x})$ denotes the median of vector $\qvec{x}$; $\qvec{x}^{\top}$ denotes transpose of vector $\qvec{x}$; $\qvec{\textit{1}}$ denotes a vector of ones; $f(.)$ denotes the probability density function; $\mathrm{mod}(.)$ denotes modulo operation; $x^{(0)}$ denotes the initial value of $x$; $\hat{\qvec{x}}$ denotes the estimate of vector $\qvec{x}$; $\Tilde{\qvec{X}}$ denotes the estimated matrix of outlier-free samples from matrix $\qvec{X}$.
\subsection{Similarity Measures for Graphs}
In graph-based clustering, a data set is represented as a weighted graph, larger weights implying higher similarity between two points. A graph $G=\{V,E,\qvec{W}\}$ is described by a set of vertices $V$, a set of edges $E$, and a weighted adjacency matrix $\qvec{W}$ that is commonly referred to as the affinity (or similarity) matrix. For a data set ${\qvec{X}\triangleq[\qvec{x}_{1},\qvec{x}_{2},\dots,\qvec{x}_{n}] \in \mathbb{R}^{m\times n}}$ with $m$ features and $n$ observations, the associated affinity matrix $\qvec{W}\in \mathbb{R}^{n\times n}$ displays the similarity between pairs of feature vectors. A popular choice, assuming non-negative and symmetrical weight functions is the cosine similarity\cite{NetworksNewman}
\begin{equation}\label{eq:cosinesim}
w_{i,j}=w_{j,i}=\frac{\qvec{x}_i^\top\qvec{x}_j}{\|\qvec{x}_i\|_2\|\qvec{x}_j\|_2},\hspace{3mm}i,j=1,\dots,n
\end{equation}
where $w_{i,j}$ is the similarity coefficient of affinity matrix $\qvec{W}$ for the $i$th and $j$th data vectors $\qvec{x}_i$ and $\qvec{x}_j$, respectively. Alternatively, a graph can be formed, for example, by using Pearson's linear correlation coefficient, as
\begin{equation}\label{eq:Pearson}
    w_{i,j}=\frac{({\qvec{x}}_i -\hat{\mu}_i)^\top({\qvec{x}}_j -\hat{\mu}_j)}{\hat{\sigma}_i\hat{\sigma}_j}
  \end{equation}  
with associated sample means $\hat{\mu}_i$, $\hat{\mu}_j$, and sample standard deviations $\hat{\sigma}_i$, $\hat{\sigma}_j$, respectively for $i,j=1,\dots,n$. In real-world problems with large and noisy data sets, directly inferring clusters from Eq.~\eqref{eq:cosinesim} or \eqref{eq:Pearson} may be inefficient, or even infeasible. For such settings, sparse graphs provide a more suitable approach to unveil the underlying data structure. It has been shown that $\ell_1$ graphs can provide a linear sparse representation of the data with respect to an overcomplete dictionary of basis elements \cite{l1graph}. If the solution is sufficiently sparse \cite{SparseEnough}, the problem can be generalized over all atoms in $\qvec{X}$, using the data matrix itself as dictionary\cite{SparseSubspace}, i.e.,
\begin{equation}\label{eq:SparseRep}
   \hat{\qvec{A}}=\arg\min\|\qvec{A}\|_1 \hspace{2mm} s.t. \hspace{0.3cm} \qvec{X}=\qvec{X}\qvec{A}, \hspace{2mm} \mathrm{diag}(\qvec{A})=0,
\end{equation}
where $\hat{\qvec{A}}\in \mathbb{R}^{n\times n}$ denotes the estimated coefficient matrix and $\mathrm{diag}(\qvec{A}) \in \mathbb{R}^{n}$ is the vector of diagonal elements of coefficient matrix $\qvec{A}$. 


\vspace{-3mm}
\subsection{Problem Statement}
Given a data set ${\qvec{X} \in \mathbb{R}^{m\times n}}$, the aim of this work is to find a label vector $\mathbf{c}_K \in \mathbb{R}^{n}$ that partitions ${\qvec{X}}$ into $K$ independent and mutually exclusive communities.  The true community number $K$ is assumed to be unknown. Further, we assume that ${\qvec{X} \in \mathbb{R}^{m\times n}}$ may be subject to heavy-tailed noise and outliers, which obscure the data structure. Computational efficiency is also of practical interest. \emph{Summarizing, the overall aim is to develop a fast and robust clustering algorithm based on computing a sparse graph model}.



\section{Sparsity-aware Robust Community Detection (SPARCODE)}\label{sec:Proposed}
\begin{figure}
 \centering
 \resizebox{\columnwidth}{!}{ \input{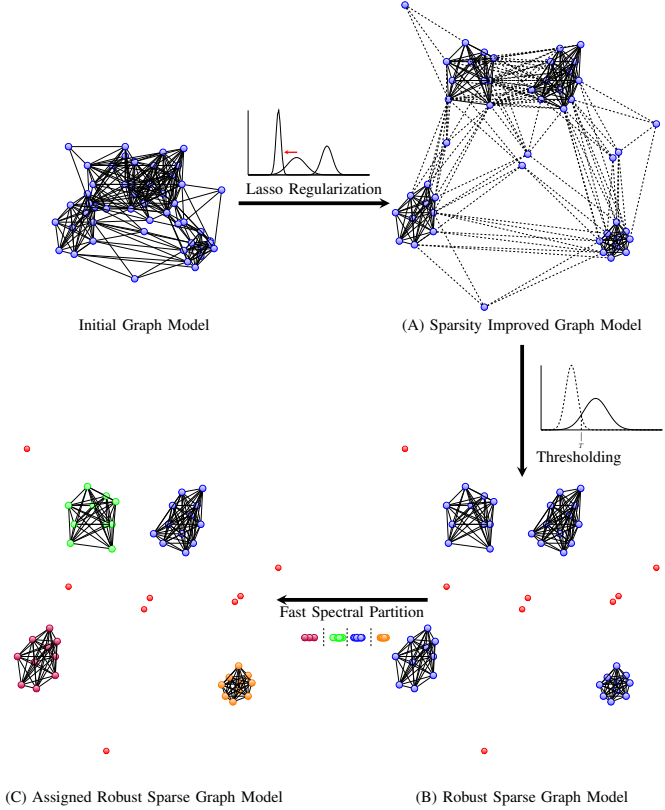}
 }
 \caption{The main steps of sparsity-aware community detection (SPARCODE).}
 \label{fig:mainideas}
\end{figure}

The main ideas of the proposed method are briefly summarized to provide a general understanding before going into the details. A high-level flow diagram to illustrate its key steps is provided in Fig.~\ref{fig:mainideas}. The community detection problem is addressed as spectral partitioning of sparse graphs. The proposed algorithm, which we call Sparsity-aware Robust Community Detection (SPARCODE), starts with a densely connected weighted graph. Assuming that edges within the communities are more densely connected than the remaining edges, the first step is to increase the sparsity of a given graph by pushing towards zero the similarity coefficients that belong to undesired and negligible edges. The sparsity improved  graph model is obtained via an $\ell_1$-penalized precision matrix estimation. The penalty parameter is optimized to provide a mapping of the coordinates from different clusters in such a way that they are embedded as far as possible on the real line. In particular, we split the Fiedler vector and optimize according to what we call the polarization score. Then, undesired and negligible edges are removed from the sparsity improved graph model by applying a threshold, and the graph construction is performed in a robust manner. The outliers, which are represented as red points in Fig.~\ref{fig:mainideas}, are detected based on connectivity of nodes in the improved sparse graph model. Finally,  fast spectral partitioning is performed on the outlier-free vertices of the robust sparse graph model by mapping each vertex onto a line and applying a balanced partition. The number of communities is estimated using modularity optimization on the partitions. The steps are detailed in the following sections, a pseudocode summary is given in Algorithm~1.

\subsection{Compute Sparsity Improved Graph Model}

It is well known (see, e.g. \cite{neighborselect}) that recovering the structure of an undirected Gaussian graph is equivalent to the recovery of the support of the precision matrix $\boldsymbol{\Theta}$. If the $i,j$th entry of  $\boldsymbol{\Theta}$ equals zero, the two corresponding variables are conditionally independent. Therefore, a sparse precision matrix leads to a sparsely connected graph. A very popular approach to estimate a sparse precision matrix is the graphical LASSO \cite{Glasso}, which maximizes a penalized Gaussian likelihood 
\begin{equation}\label{eq:Glasso}
\hat{\boldsymbol{\Theta}}=\underset{\hat{\boldsymbol{\Theta}}\in \mathbb{S}^{o}_{++}}{\arg\max}\{\mathrm{log}|\boldsymbol{\Theta}|-\mathrm{tr}(\qvec{S}\boldsymbol{\Theta})-\rho\|\boldsymbol{\Theta}\|_1\},
\end{equation}
where $\mathrm{tr}$ denotes trace, $\qvec{S}\in \mathbb{R}^{n\times n}$ is the sample covariance matrix of $\qvec{W}$ and $\rho$ is a sparsity inducing penalty parameter. According to Eq.~\eqref{eq:Glasso}, $\hat{\boldsymbol{\Theta}}$ is nonnegative definite. However,  graphical LASSO estimates, do generally not satisfy $\hat{\boldsymbol{\Theta}}\in \mathbb{S}^{o}_{++}$ \cite{glassolimitation}. This property is not crucial in the SPARCODE method, because the sign of the coefficients in the similarity matrix is not relevant for reconstructing edges in the sparsity improved graph model. Therefore, a sparsely connected graph ${\hat{G}=\{V,\hat{E},\hat{\qvec{W}}\}}$ is obtained by using the (element-wise) absolute value of the estimated inverse covariance matrix $\hat{\boldsymbol{\Theta}}$ as affinity matrix, i.e. $\hat{\qvec{W}}=|\hat{\boldsymbol{\Theta}}|$. 

While graphical LASSO provides an efficient way to compute a solution to Eq.~\eqref{eq:Glasso}, the resulting graph structure critically depends on the value of $\rho$. A key contribution of this work is to provide a simple but effective clustering-oriented answer to how to optimize $\rho$. The intuition behind our approach is to determine a value for $\rho$ by mapping the coordinates from different communities in such a way that they lie as far as possible on the real line. 

According to \cite{Geometryflows}, the mapping of vertices onto the real line by minimizing the sum of the edge lengths while maintaining an average unit Euclidean distance between random pairs of points may lead to an excellent partition by cutting the line at a random point. Achieving such a mapping is NP hard, though there are several approximations in literature that achieve the sparsest cut  \cite{LeightonRao,AppSparsestcut1,AppSparsestcut2}. SPARCODE relies on spectral partitioning methods, which use a relaxation to map graph vertices onto the real line, providing connected points stay as close together as possible using squared Euclidean distance \cite{Laplacianeigenmaps}. Let  ${\rho_k\in\{\rho_{\mathrm{min}},\dots,\rho_{\mathrm{max}}\}}$ denote the $k$th candidate penalty parameter in Eq.~\eqref{eq:Glasso}. Then, the embedding result $\hat{\qvec{y}}_k\in\mathbb{R}^n$ of the $k$th candidate penalty parameter $\rho_k$ can be approximated by minimizing the following objective function as in \cite{Laplacianeigenmaps},
\begin{equation}\label{eq:Lapembedding}
\hat{\qvec{y}}_k=\underset{\qvec{y}_k}{\arg\min}\frac{1}{2}\sum_{i,j}\big\|y_i-y_j\big\|_2^2\hat{w}_{i,j}(\rho_k)\hspace{0.9mm} s.t.\hspace{0.9mm}
\begin{aligned}
  \qvec{y}^{\mathrm{T}}\qvec{D}_k\qvec{y}=1\\
  \qvec{y}^{\mathrm{T}}\qvec{D}_k\qvec{1}=0,        \end{aligned}
\end{equation}
where $\qvec{D}_k\in \mathbb{R}^{n\times n}$ is a diagonal weight matrix of $\hat{\qvec{W}}(\rho_k)$ for a given $\rho_k$, with weights $d_{i,i}=\sum_j\hat{w}_{i,j}(\rho_k)$ on the diagonal. The vector estimate $\hat{\qvec{y}}_k=(\hat{y}_1,\ldots,\hat{y}_n)^\top$ is known as the Fiedler vector and it shows the algebraic connectivity of a graph \cite{Fiedler}.  The graph can be partitioned into two subsets by splitting the Fiedler vector, such that $\hat{y}_i\in\hat{\qvec{y}}_{k,1}$ for $\hat{y}_i\leq0$ and $\hat{y}_i\in\hat{\qvec{y}}_{k,2}$ otherwise. Here, $\hat{\qvec{y}}_{k,m}$ denotes the $m$th subset of $\hat{\qvec{y}}_{k}$ with $m=1,2$ and $i=1,\ldots,n$ \cite{Fiedlercut}. 

Based on the Fiedler vector, we propose to measure the polarization score for each candidate $\rho_k$ by evaluating
\begin{equation}\label{eq:Polarscore}
P_{\mathrm{sc}}(\rho_k)=\mathrm{med}(\hat{\qvec{y}}_{k,1})-\mathrm{med}(\hat{\qvec{y}}_{k,2}),
\end{equation}
where $\mathrm{med}(\hat{\qvec{y}}_{k,m})$ denotes the median of the $m$th subset for $m=1,2$. The median is used as a robust location estimate \cite{zoubir-2018, maronna-2019} for each subset $\hat{\qvec{y}}_{k,m}$. Given a set of candidate penalty parameters ${\rho_k\in\{\rho_{\mathrm{min}},\dots,\rho_{\mathrm{max}}\}}$, we  estimate $\rho$ by maximizing the polarization score as
\begin{equation}\label{eq:MaxPolarscore}
   \hat{\rho}=\underset{\rho=\rho_{\mathrm{max}},\dots,\rho_{\mathrm{min}}}{\arg\max} S(\rho),
\end{equation}
where $\hat{\rho}$ denotes the estimated penalty parameter that provides maximum polarization in the given space. To reduce computational complexity, we first evaluate Eq.~\eqref{eq:MaxPolarscore} on a coarse grid and then use cubic spline interpolation between $\hat{\rho}$ and its neighboring samples to find the final value $\hat{\rho}$. By means of Eq.~\eqref{eq:MaxPolarscore}, we propose a problem-dependent tuning of the level of sparsity in Eq.~\eqref{eq:Glasso}. This provides us with an initial robust sparse graph model that we will improve in the following step.

\subsection{Compute a Robust Sparse Graph Model}
 By maximizing the polarization score via Eq.~\eqref{eq:MaxPolarscore}, we estimated a graph model $\hat{G}=\{V, \hat{E},\hat{\qvec{W}}(\hat{\rho})\}$, for which the edges between different communities are large, while the edges within the community are small. However, the true graph structure may still be hidden due to noise and outliers. It is intuitively clear that sparse outlying entries and noise have fewer non-zero coefficients in the affinity matrix compared to typical data points, and the value of these coefficients is negligibly small for noise \cite{SparseSubspace}. SPARCODE builds upon this graph property to detect outliers and noisy feature vectors by analyzing the connectivity of graph nodes. 

 Let $\hat{\qvec{W}}^{(s)}(\hat{\rho})=[\hat{\qvec{w}}_1^{(s)}(\hat{\rho}),\dots,\hat{\qvec{w}}_n^{(s)}(\hat{\rho})]\in\mathbb{R}^{n\times n} $ be the set of column-wise sorted similarity vectors whose $j$th element denotes the sorted $j$th similarity vector of the estimated affinity matrix $\hat{\qvec{W}}(\hat{\rho})$. Then, we obtain the aggregated set of similarity coefficients $\qvec{u}=\{u_1,u_2,\dots,u_n\}\in\mathbb{R}^{n}$ by computing 
 \begin{equation}
 \label{eq:aggreg-sim-coeff}
 u_i=\frac{1}{n}\sum_{j=1}^{n}\hat{w}^{(s)}_{i,j}(\hat{\rho}), \quad i=1,\ldots, n,
 \end{equation}
 \vspace{-1mm}
 where $u_i$ denotes the $i$th element of $\qvec{u}$ and $\hat{w}^{(s)}_{i,j}(\hat{\rho})$ is the element of $\hat{\qvec{W}}^{(s)}(\hat{\rho})$ that belongs to the $i$th row and $j$th column. From Eq.~\eqref{eq:aggreg-sim-coeff}, 
a two-mode Gaussian mixture distribution of the $i$th similarity coefficient $u_i$ can be written as
\begin{equation} \label{eq:gaussianmixture}
    f(u_i|\boldsymbol{\lambda})=\sum_{l=1}^{2}\tau_lg(u_i;\mu_l, \Sigma_l)
\end{equation}
\vspace{-1mm}
where $\boldsymbol{\lambda}=\{\boldsymbol{\mu}_l,\boldsymbol{\Sigma}_l \}$ denotes the parameter set of the model for $l=1,2$ and $g(u_i;\mu_l, \Sigma_l)$ is the $r$-variate Gaussian probability density function with mean $\mu_l$ and covariance $\Sigma_l$, and $\tau_l$ is the mixing coefficient of the $l$th subset. The probability estimate that $u_i$ belongs to the $l$th subset, with $i=1,2,\dots,n$ and $l=1,2$, can be calculated as
\begin{equation} \label{eq:posteriorprobab}
 \hat{\upsilon}_{i,l}=\frac{\tau_lg(u_i;\mu_l, \Sigma_l)}{\sum_{j=1}^{2}\tau_jg(u_i;\mu_j, \Sigma_j)}
\end{equation}
\vspace{-1mm}
\vspace{1mm}\\where $\hat{\upsilon}_{i,l}$ denotes the probability estimate of $u_i$ that belongs to the $l$th subset. Then, the threshold can be simply evaluated as the interconnection point of the two distributions, i.e.
\begin{equation} \label{eq:Threshold}
 T=\underset{u_i}{\arg\min} |\hat{\upsilon}_{i,2}-\hat{\upsilon}_{i,1}|.
\end{equation}
\vspace{-1mm}
The effect of applying the threshold defined in Eq.~\eqref{eq:Threshold}, is to remove the edges from the estimated graph model $\hat{G}=\{V, \hat{E},\hat{\qvec{W}}(\hat{\rho})\}$ so that the feature vectors in data matrix $\qvec{X}$ which have fewer nonzero coefficients with negligibly small values will have a zero degree in the graph. In this way, outliers and noisy feature vectors are detected and removed, resulting in a cleaned data matrix $\Tilde{\qvec{X}}\in\mathbb{R}^{\tilde{n}}$ and a graph $\Tilde{G}=\{\Tilde{V}, \Tilde{E},\Tilde{\qvec{W}}(\hat{\rho})\}$.

\subsection{A Fast Spectral Partition based on the Robust and Sparse Graph}\label{sec:fastspectralpartition}
Assuming that $\Tilde{G}=\{\Tilde{V}, \Tilde{E},\Tilde{\qvec{W}}(\hat{\rho})\}$ is cleaned from outliers and that it is sufficiently sparse, the graph can be partitioned into $K$ communities based on the Fiedler vector with a fast partitioning method. In practice, however, $K$ is unknown and must be estimated. We therefore present an approach to estimate the range $K_{\mathrm{min}}\leq K \leq K_{\mathrm{max}}$ using typical degrees in the graph. 


Let $\qvec{d}\in \mathbb{Z}^{+\Tilde{n}}$ denote a degree vector whose $i$th element corresponds to the degree of the $i$th feature vector in $\Tilde{\qvec{X}}\in\mathbb{R}^{\tilde{n}}$. Further, let $\qvec{p}\in \mathbb{R}^{\Tilde{n}}$ be the empirical probability of occurrence of these degrees in $\qvec{d}$. Finally, let $\qvec{h}\in \mathbb{R}^{\Tilde{n}/2-1}$ denotes the vector of degrees, whose probability is greater than the median of probabilities, i.e. $\mathrm{med}(\qvec{p})$. 
Now, the minimum and maximum number of communities $\{K_{\mathrm{min}},K_{\mathrm{max}}\}\in\mathbb{Z}^{+}$can be estimated as 
\begin{equation}\label{eq:Kmin_Kmax}
\hat{K}_{\mathrm{min}}\approx\frac{\Tilde{n}}{h_{\mathrm{max}}+1} \ \mathrm{ and } \ \hat{K}_{\mathrm{max}}\approx\frac{\Tilde{n}}{h_{\mathrm{min}}+1}
\end{equation}\vspace{1mm}\\ 
where $h_{\mathrm{min}}+1$ and $h_{\mathrm{max}}+1$ represent the minimum and maximum number of nodes in each community based on $h_{\mathrm{min}}=\mathrm{min}\{\qvec{h}\}$ and $h_{\mathrm{max}}=\mathrm{max}\{\qvec{h}\}$, respectively. The intuition underlying Eq.~\eqref{eq:Kmin_Kmax} is to define a range for a candidate number of communities from the typical connectedness of the graph.


Now, for a set of estimated candidate communities ${\hat{K}_{\mathrm{cand}} \in \{\hat{K}_{\mathrm{min}},\dots,\hat{K}_{\mathrm{max}}\}}$, the graph can be partitioned using the sorted Fiedler vector $\Tilde{\qvec{y}}^{(s)}\in\mathbb{R}^{\Tilde{n}}$ of $\Tilde{G}=\{\Tilde{V}, \Tilde{E},\Tilde{\qvec{W}}(\hat{\rho})\}$ as follows. To split the graph into an initial balanced partitioning, where each community has the same number of members, we first compute $q_{\mathrm{cand}} =\mathrm{mod}(\Tilde{n},\hat{K}_{\mathrm{cand}})$. We identify a number $q_{\mathrm{cand}}$ of mapping results that we ignore in the initial balanced partitioning. 
These are the ones for which the standard deviation of the set containing its two immediate neighbors in the sorted Fielder vector is largest.
\begin{algorithm}[tbp!]
\setstretch{1}
\DontPrintSemicolon
\KwIn{
An affinity matrix $\qvec{W}\in \mathbb{R}^{n\times n}$ }
\textbf{Step 1: Robust Sparse Graph Model}\\
\textbf{Step 1.1: Sparsity Improvement}\\
\textbf{Initialization:}\\
\For{$\rho_k^{(0)}=\rho_{\mathrm{min}}^{(0)},\dots,\rho_{\mathrm{max}}^{(0)} $}{
Estimate $\hat{\qvec{W}}(\rho^{(0)})$ via Eq.~\eqref{eq:Glasso}\\
Map each vertice onto a line using Eq.~\eqref{eq:Lapembedding}\\
Get the Fiedler vector $\hat{\qvec{y}}_k^{(0)}$\\
Split $\hat{\qvec{y}}_k^{(0)}$ into two subsets for\hspace{0.7mm}a\hspace{0.7mm}splitting value\hspace{0.7mm}$s=0$\\
Calculate\hspace{0.7mm}the\hspace{0.7mm}polarization\hspace{0.7mm}score\hspace{0.6mm}$P_{\mathrm{sc}}(\rho_k^{(0)})$\hspace{0.7mm}via\hspace{0.9mm}Eq.~\eqref{eq:Polarscore}\\
Stack $P_{\mathrm{sc}}(\rho_k^{(0)})$ into $\qvec{p}_{\mathrm{sc}}^{(0)}\in \mathbb{R}^{N_{\rho^{(0)}}}$
}
\textbf{Penalty Parameter Selection:}\\
Find $\rho_k^{(0)}$ which maximizes Eq.~\eqref{eq:MaxPolarscore} for an initial set\\
Regenerate a penalty parameter set ${\rho_k=\rho_{\mathrm{min}},\dots,\rho_{\mathrm{max}}}$ over equally spaced $N_{\rho}$ samples\\ 
Apply the same framework as in the initialization step\\
Obtain $\qvec{p}_{\mathrm{sc}}\in \mathbb{R}^{N_{\rho}}$ for ${\rho_k=\rho_{\mathrm{min}},\dots,\rho_{\mathrm{max}}}$\\
Apply cubic spline interpolation to obtain $\hat{\rho}$\\
\textbf{Step 1.2: Robustness and Outlier Detection}\\
Create $\hat{\qvec{W}}^{(s)}(\hat{\rho})\in\mathbb{R}^{n \times n}$ over a set of sorted similarity\\vectors from $\hat{\qvec{W}}(\hat{\rho})$\\
Get $\qvec{u}\in\mathbb{R}^{n}$ via Eq.~\eqref{eq:aggreg-sim-coeff}\\
Estimate $\hat{\upsilon}_{i,l}$ for each coefficient where $i=1,2,\dots,n$\\and $l=1,2$\\Calculate $T$ using Eq.~\eqref{eq:Threshold}\\Cut undesired edges in $\hat{G}$ using $T$\\Reject outliers whose degree equals zero, $d=0$\\
Form $\Tilde{\qvec{W}}(\hat{\rho})\in\mathbb{R}^{\Tilde{n}\times\Tilde{n}}$ 
over estimated outlier-free vectors\\
\textbf{Step 2: Fast Spectral Partition}\\
Estimate $\hat{K}_{\mathrm{cand}} \in \{\hat{K}_{\mathrm{min}},\dots,\hat{K}_{\mathrm{max}}\}$ using Eq.~\eqref{eq:Kmin_Kmax}\\
\For{$\hat{K}_{\mathrm{cand}} =\hat{K}_{\mathrm{min}},\dots,\hat{K}_{\mathrm{max}}$}{
Create $\Tilde{\qvec{y}}\in\mathbb{R}^{\Tilde{n}}$ using Eq.~\eqref{eq:Lapembedding}\\
Calculate $q_{\mathrm{cand}}$ as $q_{\mathrm{cand}}=\mathrm{mod}(\Tilde{n},\hat{K}_{\mathrm{cand}})$\\
Define initially ignored mapping results in $\Tilde{\qvec{y}}^{(s)}$\\
Apply an initial partition on $\Tilde{\qvec{y}}^{(s)}\in \mathbb{R}^{\Tilde{n}-q_{\mathrm{cand}}}$\\
Assign initially ignored mapping results\\
Form $\Tilde{\qvec{c}}_{\mathrm{cand}} \in \mathbb{R}^{\Tilde{n}}$ for $\hat{K}_{\mathrm{cand}}$\\
Calculate $Q_{\hat{K}_{\mathrm{cand}}}$ via Eq.~\eqref{eq:modscore}\\
Stack $Q_{\hat{K}_{\mathrm{cand}}}$ into $\qvec{q}\in\mathbb{R}^{N_K}$ 
}
Estimate $\hat{K}$ using Eq.~\eqref{eq:maxmodscore}\\

\caption{SPARCODE\label{IR}}
\KwOut{A vector $\Tilde{\qvec{c}}$ for $\hat{K}$}
\end{algorithm}
Then, using the remaining mapping results, the initial partition can be realized by splitting the remaining mapping results into $\hat{K}_{\mathrm{cand}} $ equal communities. 
 Finally, the residual $q_{\mathrm{cand}}$ mapping results can be assigned by minimizing the distance to the center of the initially defined communities on the line. Now, the estimated label vector of a given candidate number of community $\Tilde{\qvec{c}}_{\hat{K}_{\mathrm{cand}}}\in\mathbb{R}^{\Tilde{n}}$ can be computed for all feature vectors of $\Tilde{\qvec{X}}$. 

Finally, the number of communities $K$ can be estimated by comparing in terms of their modularity, the quality of different partitions of the robust graph model i.e.
\begin{equation}\label{eq:maxmodscore}
\hat{K}=\underset{\hat{K}_{\mathrm{cand}} =\hat{K}_{\mathrm{min}},\dots,\hat{K}_{\mathrm{max}}}{\arg\max} \{Q_{\hat{K}_{\mathrm{cand}}}\},
\end{equation}\vspace{1mm}\\
where
\begin{equation}\label{eq:modscore}
Q_{\hat{K}_{\mathrm{cand}} }=\frac{1}{2g}\sum_{j,m}^{\tilde{n}}\Big[{\Tilde{w}}_{j,m}-\frac{\Tilde{v}_j\Tilde{v}_m}{2g}\Big]\delta(\Tilde{c}_j,\Tilde{c}_m)
\end{equation}\vspace{1mm}\\
is the modularity score \cite{Newman2002, Newman2006} for the $i$th candidate number of communities, $\Tilde{w}_{j,m}$ denotes the weight of the edge between the $j$th and $m$th feature vector of $\Tilde{\qvec{X}}$, $\Tilde{v}_j=\sum_{m}\Tilde{w}_{j,m}$ is the sum of the weights of edges attached to vertex $j$, $\Tilde{c}_j$ is the estimated community label associated to vertex $j$, $g=\frac{1}{2}\sum_{j,m}\Tilde{w}_{j,m}$ and the function $\delta(\Tilde{c}_j,\Tilde{c}_m)$ equals 1 if $\Tilde{c}_j=\Tilde{c}_m$ and is zero otherwise. 

\vspace{-2mm}
\section{Experimental Evaluation}\label{sec:Experimental}
In this section, the community detection performance of SPARCODE is benchmarked against state-of-the-art community detection algorithms. We consider a variety of clustering and graph partitioning data sets to demonstrate the applicability of SPARCODE in both of these community detection settings. We select as competitors the following community detection algorithms: Newman's Greedy Algorithm (NGA) \cite{NGA}, Le Martelot \cite{Martelot}, Newman's eigenvector method \cite{NE}, singular value decomposition-based community detection (SVD) \cite{SVD}, the Louvain method \cite{Louvain}, the Bayesian approach (BC) \cite{Bayesian} and the Combo method \cite{Combo}. The SVD method is applicable only on bipartite graphs, for details see \cite{SVD}. Bayesian cluster enumeration (BCE) \cite{BayesianTekle}, dip-means and kernel dip-means (K. dip-means)\cite{dipmeans}, x-means \cite{xmeans}, Gaussian k-means (g-means) \cite{gmeans} and density based spatial clustering (DBSCAN) \cite{DBSCAN} are used as cluster-based competing approaches. The performance measures are evaluated both on synthetic and real-world data sets. 

All SPARCODE implementations use the same default parameters as follows: 
$\rho_{\mathrm{min}}^{(0)}=0.1$, $\rho_{\mathrm{max}}^{(0)}=0.99$, 
$N_{\rho}^{(0)}=N_{\rho}=5$. A MATLAB code for SPARCODE is available at:\\
\vspace{5mm}
\url{https://github/SPARCODE}
\subsection{Performance Measures}
The empirical probability of detection $p_{\mathrm{det}}$, the modularity score $Q$ and the computation time $t$ are used as performance measures. The empirical probability of detection is estimated as
\begin{equation}\label{eq:probdetection}
p_{\mathrm{det}}=\frac{1}{N_E}\sum_{i=1}^{N_E}\mathbbm{1_{\{\hat{K}=K\}}},
\end{equation}
where $N_E$ denotes the total number of performed experiments, $\hat{K}$ is the estimated number of communities and $\mathbbm{1_{\{\hat{K}=K\}}}$ is the indicator function that is defined as
\begin{equation}\label{eq:indicatorfunc}
\mathbbm{1_{\{\hat{K}=K\}}}=\begin{cases}
     1 ,        & \text{if } \hat{K}=K\\
     0 ,           & \text{otherwise}
\end{cases}.
\end{equation}
The modularity score of the estimated community number $Q_{\hat{K}}$ can be calculated using Eq.~\eqref{eq:modscore}. The modularity score of SPARCODE is computed based on the affinity matrix of the sparsity improved robust graph model.


\subsection{Synthetic Graph Model 1: Correlated Communities Study with $K$=7}
An undirected weighted graph model is created, consisting of seven communities with different sample sizes for each community. The first six communities inhibit correlation between communities in addition to correlations within the community. In contrast, the seventh community only has correlations within the community. Moreover, the density of the seventh community is assumed to be higher compared to that of the other six. For illustration purposes, the graph for the designed affinity matrix is shown in Fig.~\ref{fig:scenario1a}, all parameters to generate the data set are given in Appendix~\ref{app:Scenario1}. 

Tab.~\ref{tab:tableScenario1} summarizes the community detection performance results for Scenario~1 in terms of the estimated community number, the modularity score and the computation time. The considered setting is challenging for most algorithms, with the exception of that in \cite{Bayesian}, and all benchmark community detection algorithms underestimate the number of clusters due to the correlations between communities. SPARCODE correctly finds the number of communities and outperforms the method in \cite{Bayesian} both in terms of the modularity score and the computation time.   As shown in Fig.~\ref{fig:scenario1b}, the estimated sparsity improved robust graph model clearly partitions the network into seven communities. 

\begin{figure}[!tbp]
  \centering
  \subfloat[Initial graph model]
  {\includegraphics[trim={25mm 9mm 25mm 20mm},clip,width=4.35cm]{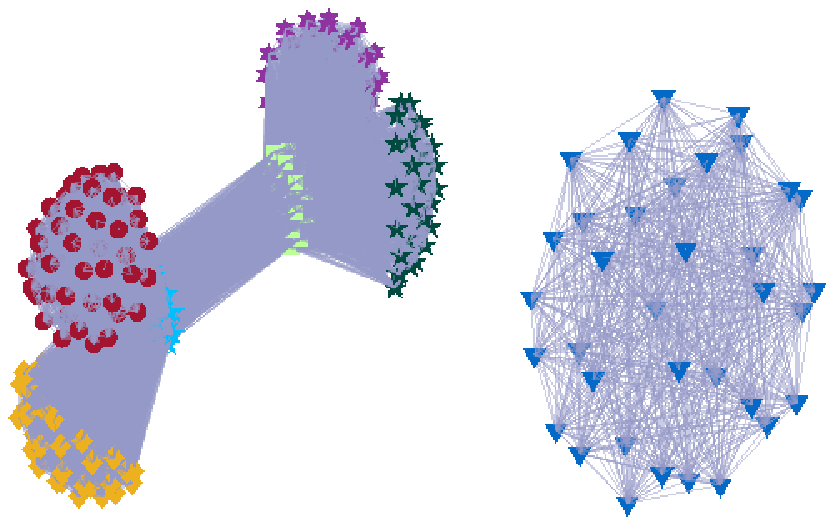}\label{fig:scenario1a}}
  \subfloat[Estimated graph model]
  {\includegraphics[trim={25mm 9mm 25mm 20mm},clip,width=4.35cm]{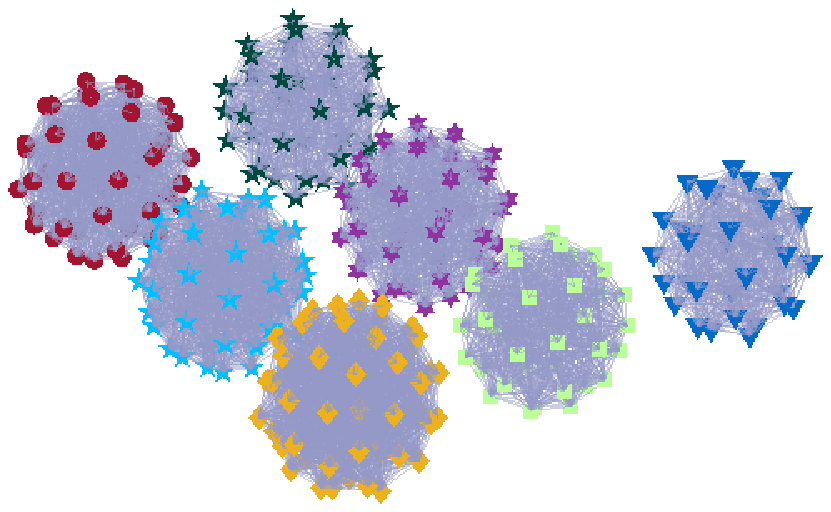}\label{fig:scenario1b}}\\
  \label{fig:scenario1graph}
  \caption{Graphical models of Scenario~1.}
\end{figure}
\begin{table}[t!]
\centering
\begin{tabular}{ p{2cm}p{1.25cm}p{1.25cm}p{1.25cm} }
\hline\hline
\toprule
Method &  $\hat{K}$  &  $Q$ & $t$ \\
\hline
\midrule
NGA & 4 & 0.5524 & 404.389  \\
Martelot & 4 & 0.5763 & 0.046  \\
NE & 1 & 0 & 0.023 \\
Louvain & 5 & 0.597 & 0.023 \\
Bayesian & 7 & 0.525 & 1.91  \\
Combo & 5 & 0.597 & 0.91  \\
SPARCODE & 7 & 0.675 & 0.768  \\
\hline
\bottomrule
\end{tabular}
\caption{\label{tab:table-name}Performance of seven graph-based approaches on Scenario~1.}
\label{tab:tableScenario1}
\end{table}

\subsection{Synthetic Graph Model 2: Robustness Study with $K$=3}
\vspace{-0.7mm}
 An undirected weighted graph consisting of three communities in the presence of outliers is created. The communities are correlated with each other in addition to exhibiting strong correlations within. The outliers correlate equally with all communities with negligibly small correlation coefficients. The graph for the designed affinity matrix of Scenario~2 is shown in Fig.~\ref{fig:scenario2a}, all parameters to generate the data set are given in Appendix~\ref{app:Scenario2}. While the NE, NGA and Bayesian community detection algorithms overestimate the community number because of outliers, the remaining competitor community detection methods estimate the community number correctly. However, for the competing methods, the outliers cause a noticeable drop in modularity. By contrast, the SPARCODE algorithm, overcomes this problem by detecting these outliers in the graph model and providing robustness by minimizing the effect of outliers on the quality of the partition. Fig.~\ref{fig:scenario2b} shows based on the robust sparse graph model, that a separation of the communities is possible and outliers can easily be distinguished from other data points.
\begin{figure}[tbp!]
  \centering
  \subfloat[Initial graph model]
  {\includegraphics[trim={25mm 13mm 25mm 14.5mm},clip,width=4.35cm]{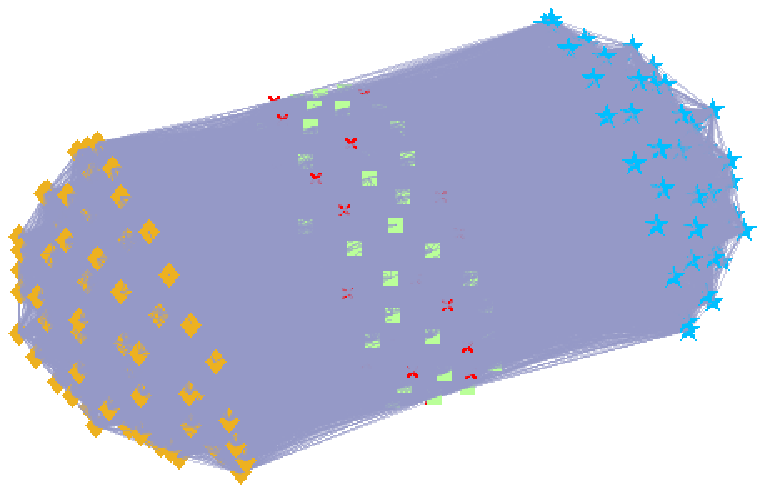}\label{fig:scenario2a}}
  \subfloat[Estimated graph model]
  {\includegraphics[trim={25mm 13mm 25mm 14.5mm},clip,width=4.35cm]{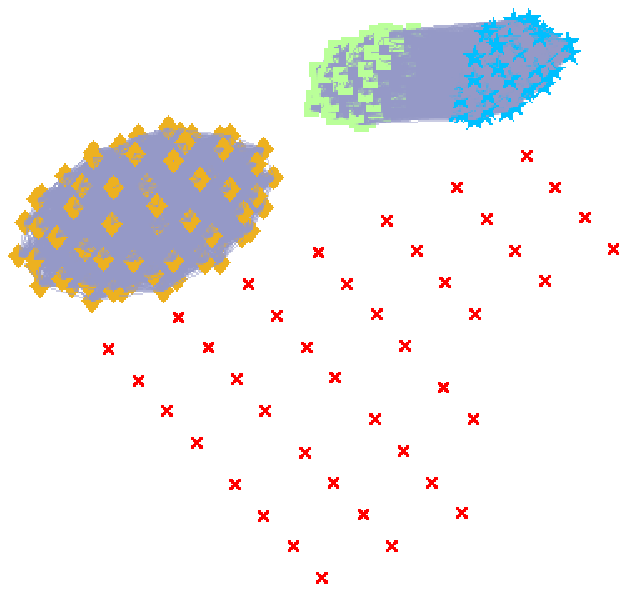}\label{fig:scenario2b}}\\
  \label{fig:scenario1graph}
  \caption{Graphical models of Scenario~2.}
\end{figure}
\vspace{-6mm}
\begin{table}[t!]
\centering
\begin{tabular}{ p{2cm}p{1.25cm}p{1.25cm}p{1.25cm} }
\hline\hline
\toprule
Method &  $\hat{K}$  &  $Q$ & $t$ \\
\hline
\midrule
NGA & 117 & 0.111 & 775.639  \\
Martelot & 3 & 0.326 & 0.054 \\
NE & 4 & 0.288 & 0.036 \\
Louvain & 3 & 0.326 & 0.041 \\
Bayesian & 10 & 0.3232 & 6.879  \\
Combo & 3 & 0.325 & 0.88  \\
SPARCODE & 3 & 0.496 & 0.753 \\
\hline
\bottomrule
\end{tabular}
\caption{\label{tab:table-name}Performance of seven graph-based approaches on Scenario~2.}
\label{tab:tableScenario2}
\end{table}
\subsection{Executing Time}
\begin{figure}[tbp!]
 \centering
  
  \subfloat[Executing time for growing number of nodes (zoom on small networks).\label{fig:allmethodscomptimesmall}]{
  \includegraphics[trim={2mm 0mm 0 0mm},clip,width=9.3cm,height=7.5cm]{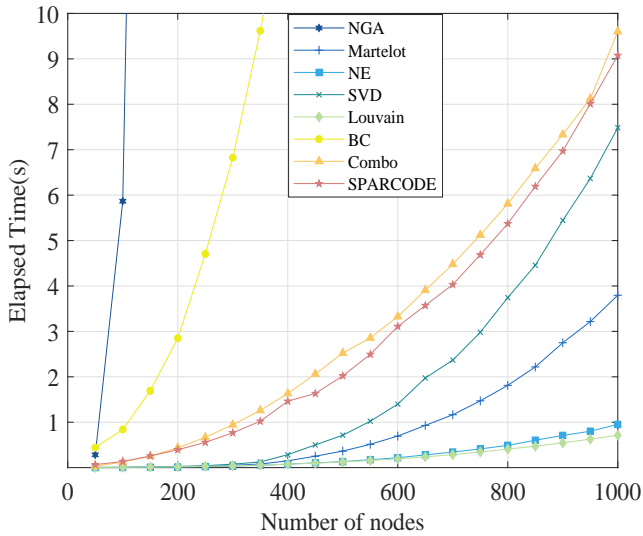}
  }
  \vspace{5mm}
    \subfloat[Executing time for growing number of nodes.\label{fig:allmethodscomptime}]{
  \includegraphics[trim={1mm 0mm 0 0mm},clip,width=9.3 cm,height=7.5cm]{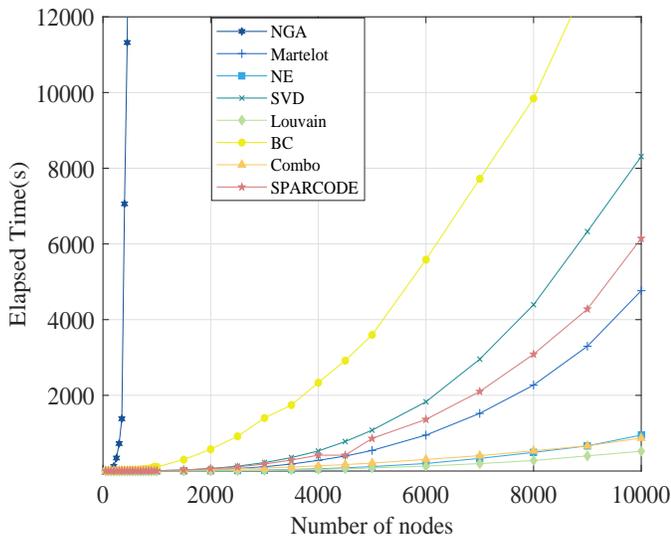}
  }
\caption{Execution time for a growing number of nodes for Scenario~2. The upper figure zooms into the region concerning networks up to a size of 1000 nodes.}
\label{fig:allmethodscomptimeboth}
\end{figure}

The executing time is reported a function of the number of nodes in the network.  All experiments are performed based on Scenario~2 that was explained in the previous section. All implementations are in MATLAB using the default parameters given by the authors, except for the Combo algorithm, for which we use the available C implementation. The results are summarized in Fig.~\ref{fig:allmethodscomptimeboth}.  For sample sizes up to 1000 nodes, SPARCODE is comparable to Combo (see Fig.~\ref{fig:allmethodscomptimesmall}), though for large networks, the executing time of SPARCODE grows quicker (see Fig.~\ref{fig:allmethodscomptime}). 

\begin{table*}
\centering
\begin{tabularx}{\textwidth}{l*{9}{>{\centering\arraybackslash}X}}
\hline\hline
 & \multicolumn{8}{c}{$\hat{K}$ for Different Community Detection Methods} &  \\
\cline{2-9}
Network & NGA & Martelot & NE & SVD & Louvain & Bayesian & Combo & SPARCODE & K\\
\midrule
Karate \cite{Karate},& 2 & 2 & 3 & 2 & 2 & 3 & 2 & 2 &2\\
Dolphins \cite{Dolphins},&2 & 3 & 6 & 2 & 4 & 2 & 4 & 2 & 2\\
Football \cite{Newman2002},&7& 4 & 8 & - & 7 & 5 & 7 & 8 &12\\
P. Blogs \cite{PoliticalBlogs},& - & 541 & 1& 1&505 & 7 & 4&2&2\\
Jazz \cite{Jazz},& 20 & 21 & 7& - & 21 & 10 & 3 & 4& 4\\
C.S.M. \cite{Carpinteria},& 10 & 12 &8 & 1 & 13 &10 &5 & 2&2,4\\
C. Elegans \cite{Celegans},& 81 & 3 & 10&- &3 & 5 & 3&3&3\\
\hline\hline
\end{tabularx}
\caption{\label{tab:table-name}Performance of graph-based approaches on well-known networks. The results whose computation take more than 12 hours and nontarget networks for SVD method denoted as "-".}
\label{tab:tablerealnetwork}
\end{table*}
\vspace{-8mm}
\subsection{Real-World Benchmark Graph Models}
The performance is benchmarked on the following seven real-world networks:
    
    \textit{Zachary's Karate Club (Karate):} The network is a social bipartite network that consists of friendship between 34 members of a karate club \cite{Karate}.
    
     \textit{Dolphins:} The network is a social network that consists of social interactions of 62 dolphins with $K=2$ communities based on the reaction after a dolphin left from the group \cite{Dolphins}.
    
     \textit{American College Football (Football):} The network represents 115 US college teams and the games that they played \cite{Newman2002}. The ground truth for the community number is 12.
     
     \textit{Political Blogs (P. Blogs):} The network consists of blogs about US politics with 1490 nodes and $K=2$ communities as "liberal" and "conservative" \cite{PoliticalBlogs}.
     
  
   \textit{Jazz Musicians (Jazz):} The network consists the collaboration of jazz musicians which can be divided into  $K=4$ communities based on cities where bands are recorded \cite{Jazz}.
   
   \textit{Carpinteria Salt Marsh (C.S.M.):} The network is a type of food web which can be divided into $K=2$ based on species as parasites and free-living or $K=4$ subwebs based on links, e.g. parasites-parasites, predator-parasites \cite{Carpinteria}.
  
    \textit{Caenorhabditis Elegans (C. Elegans):} The biological network examines the neuronal layout of C. Elegans for 279 neurons that can be partitioned into $K=3$ communities \cite{Celegans}.

The cosine similarity, defined in  Eq.~\eqref{eq:cosinesim}, is used to obtain the affinity matrix $\qvec{W}$ for all networks except for the political blogs network, which is examined with Pearson's linear correlation coefficients as designed in Eq.~\eqref{eq:Pearson} because none of the algorithms estimated the community number correctly with the cosine similarity. 

\vspace{5mm}
The community detection results are summarized in Tab.~\ref{tab:tablerealnetwork}. The results include the estimated community number for all cases for which the computation time was lower than 12 hours. The community detection results of the SVD approach are given for bipartite networks, only. As can be seen, the SPARCODE algorithm shows the best overall performance, and estimates the community number correctly for all networks, except for the Football network, for which it nevertheless provides the value that is closest to the true community number. 
\vspace{3mm}
\subsection{Real-World Radar-Based Gait Analysis Data Sets:}\label{sec:radarclusteringdata sets}
\begin{figure}[!tbp]
\begin{minipage}[b]{1.0\linewidth}
  \centering
  \includegraphics[trim={0 0mm 0 0mm},clip,width=8.5 cm]{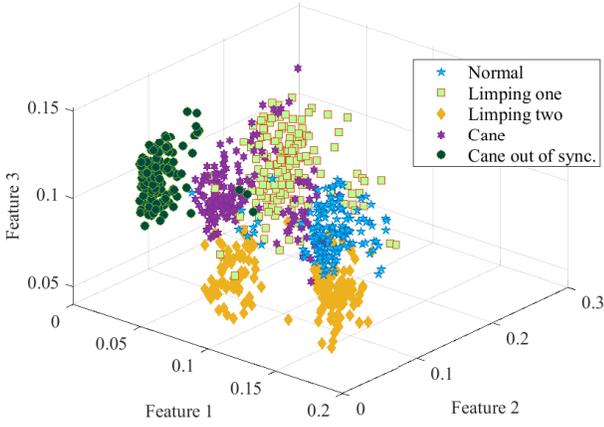}
\end{minipage}
\caption{Scatter plot for three important features of radar-based human gait data belonging to five object communities.}
\label{fig:humangaitdata}
\end{figure}

\vspace{1mm}
The performance is benchmarked on the following two radar data sets:
\vspace{2mm}

\textit{Human Gait Data Set:}
As detailed in \cite{humangait}, the experimental data was collected in an office environment at Technische Universität Darmstadt using a 24 GHz radar system \cite{radarinfo}.  The data can be grouped into $K=5$ different object communities, which are normal walk ('Normal'), limping with one leg (Limping one), limping with two legs ('Limping two'), walking with a cane ('Cane') and walking with a cane out of synchronization ('Cane out of sync.'). The data contains 16 measurements per subject (8 towards, 8 away from the radar) and ten subjects. The duration of each measurement is equal to six seconds. In total, 800 measurements for five different gait communities of ten subjects were used in our experiments. As an illustration, a scatter plot of three important features of the radar-based human gait data is shown in Fig.~\ref{fig:humangaitdata}. It can be seen that the outlying observations corresponding to the 'Cane' community overlap with the true observations of 'Normal'. Moreover, the outliers of ’Limping two’ have a considerable sample size and strong correlations with 'Normal', which makes it a challenging scenario for any community detection algorithm. 
\vspace{2mm}

\begin{figure}[!tbp]
  \centering
  \subfloat[Initial graph model]
  {\includegraphics[trim={35mm 20mm 25mm 30mm},clip,width=4.5cm]{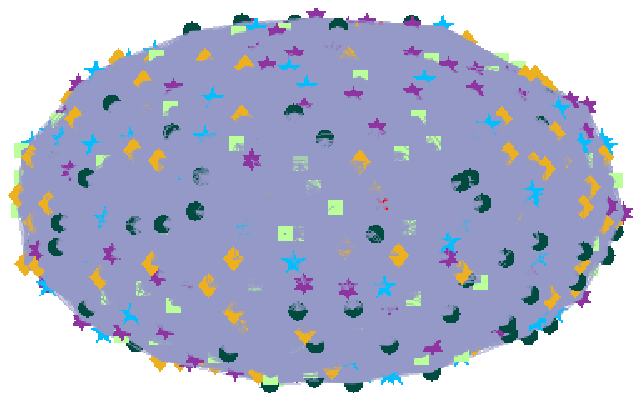}\label{fig:HumanGaitGraph1}}
  \subfloat[Estimated sparsity improved graph model]
  {\includegraphics[trim={50mm 20mm 25mm 30mm},clip,width=4.5cm]{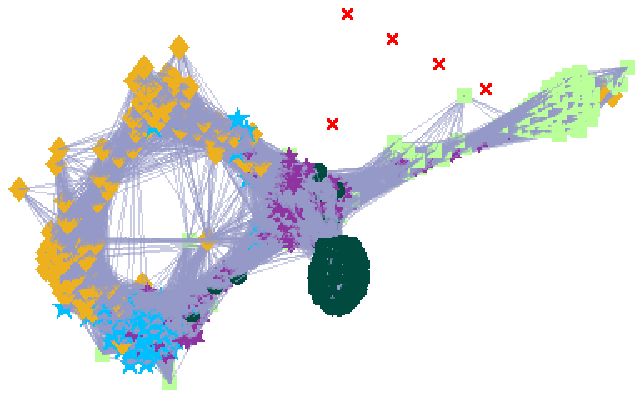}\label{fig:HumanGaitGraph2}}\\
  \caption{Graphical models of human gait data set.}
\end{figure}

In order to represent the underlying structure, for SPARCODE, similarity measures as it is explained in Eq.~\eqref{eq:SparseRep} and Eq.~\eqref{eq:Pearson} are used respectively, and the resulting graph is shown in Fig.~\ref{fig:HumanGaitGraph1}. Clearly, the separation into communities from such a graph is extremely difficult. In contrast, the sparsity improved graph model shown in Fig.~\ref{fig:HumanGaitGraph2} better reveals separated communities. 

 \textit{Person Identification Data Set:} As detailed in \cite{PersonIdentification}, the experimental data has been collected in an office environment at Technische Universität Darmstadt using the same radar system as explained for the previous data set. The data can be grouped into $K=4$ object communities, each representing a different person. The data set includes the measurements of four test subjects that are walking slowly and without swinging their arms, towards and away from the radar. The duration of recordings is equal to six seconds and the number of measurements is equal to 13 for each person. In total, 187 stride pairs that are obtained from 52 observations of the four subjects are used in community detection.
 \begin{figure}[!tbp]
\begin{minipage}[b]{1.0\linewidth}
  \centering
  \includegraphics[trim={0 0mm 0 0mm},clip,width=8.5 cm]{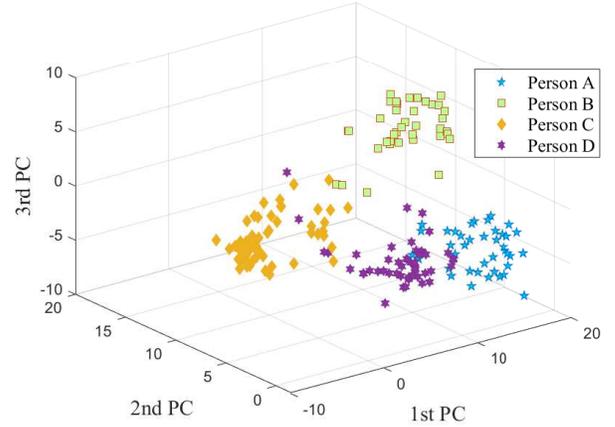}
\end{minipage}
\caption{Scatter plot for the first three principal components (PC) of person identification data over four object communities.}
\label{fig:personidentification}
\end{figure}
\begin{figure}[!tbp]
  \centering
  \subfloat[Initial graph model]
  {\includegraphics[trim={35mm 20.7mm 18mm 33.09mm},clip,width=4.35cm]{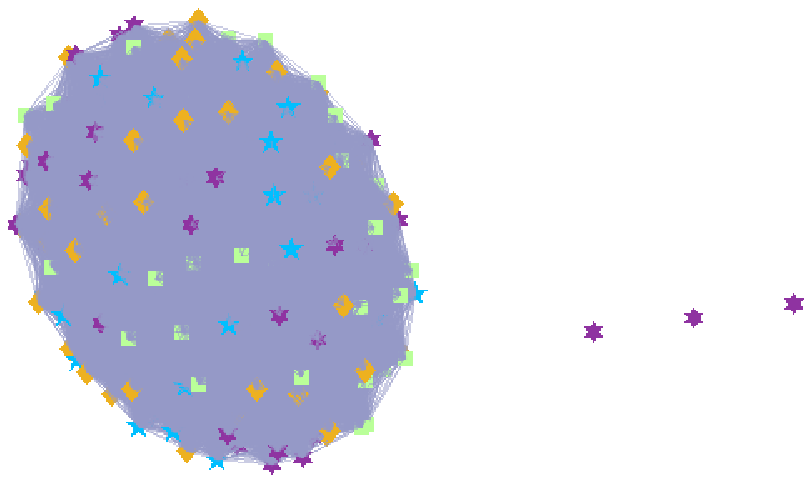}\label{fig:PersonIdenGraph1}}
  \subfloat[Estimated sparsity improved graph model]
  {\includegraphics[trim={30mm 13.7mm 30mm 29.09mm},clip,width=4.35cm]{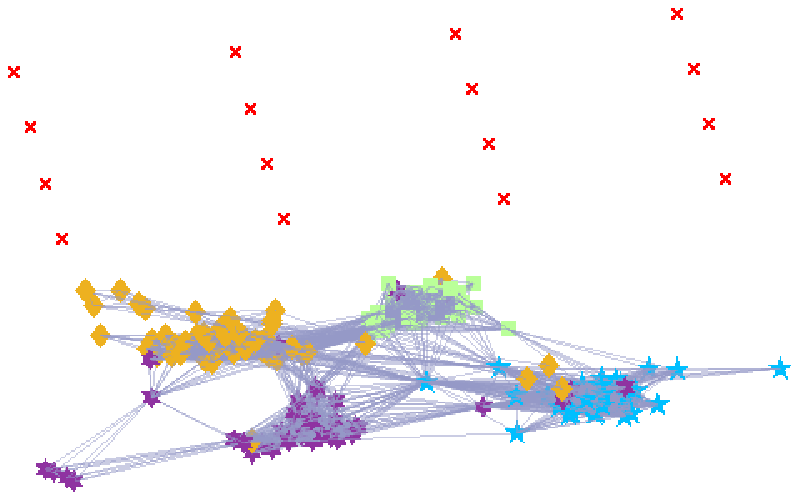}\label{fig:PersonIdenGraph2}}\\
  \caption{Graphical models of person identification data set.}
\end{figure}

The scatter plot of the person identification data set containing $K=4$ different persons is shown in Fig.~\ref{fig:personidentification}. Even though, judging based on visual impression, the examination of the person identification data set seems easier in comparison to human gait data, it still has communities that correlate with each other which makes the detection of the community number very difficult.
\begin{table*}
\centering
\begin{tabularx}{\textwidth}{l*{9}{>{\centering\arraybackslash}X}}
\hline\hline
 & \multicolumn{8}{c}{$\hat{K}$ for Different Community Detection Methods} &  \\
\cline{2-9}
Data Set & NGA & Martelot & NE & SVD & Louvain & Bayesian & Combo & SPARCODE & K\\
\midrule
Human Gait \cite{humangait},& -&3&1 & - & 3 & 5 &3 &5 &5\\
Person Identification \cite{PersonIdentification},&9&11&1&-&8&8&5 &4&4\\
Iris \cite{Iris},&1&2 &2& -&2 &2 &2&3&3\\
Ionosphere \cite{Ionosphere},& 109&106 &1 & 1 &101 &10 &6& 2&2\\
Parkinson Acoustic \cite{ParkinsonAcoustic},&1 &1&55&1&1&2&1 &2&2\\
D. Retinopathy \cite{D.Retinopathy},&-&2&2 &1 &2 &2 &2 &2&2\\
Sonar \cite{Sonar},&1 &2 &28 &1 &3 &2 & 2&2&2\\
QSAR Bioconcentration \cite{QSAR},&-&3&1&-&3&7&3&3&3
\\
Cardiotocography \cite{Cardiotocography}, &- &2 &206&- &2 &15 & 2&3&3,10\\
Divorce Predictors \cite{DivorcePredictors},&3 &3 &1&1&3&7 &2&2&2\\
\hline\hline
\end{tabularx}
\caption{\label{tab:table-name}Performance of graph-based approaches on clustering data sets. The results whose computation take more than 12 hours and nontarget networks for SVD method denoted as "-".}
\label{tab:tablerealclusters}
\end{table*}
Just like for the previous radar data set, SPARCODE uses similarity measures as defined in Eq.~\eqref{eq:SparseRep} and Eq.~\eqref{eq:Pearson}, respectively. The initial and the sparsity improved robust graph models of the person identification data set are shown in Fig.~\ref{fig:PersonIdenGraph1} and Fig.~\ref{fig:PersonIdenGraph2}, respectively. As can be seen, the estimated sparsity improved graph model is separable into four communities by eliminating the outliers are of zero degree. Therefore, a simple graph partitioning method is sufficient to partition such a robust sparse graph model into the correct number of communities. 
\vspace{-5mm}
\subsection{Real-World Cluster Analysis Benchmark Data Sets:}\label{sec:clusteringdatasets}
In this section, we benchmark the performance of SPARCODE on eight well-known data sets from the UCI Machine Learning Repository. These are:

\textit{Fisher's Iris Data Set (Iris):} The iris data set includes 150 observations from three species of the Iris flower \cite{Iris}.
  
\textit{Ionosphere Data Set:} The data set includes 351 radar returns from the ionosphere in order to define quality for further analysis \cite{Ionosphere}. The subspace number is equal to two.
  
\textit{Parkinson Acoustic Data Set:} The data set consists of replicated acoustic features of Parkinson's disease with 240 instances from two communities which are "healthy" and "patient" \cite{ParkinsonAcoustic}.
  
\textit{Diabetic Retinopathy Debrecen Data Set (D. Retinopathy):} The data set includes image-based features of diabetic retinopathy with 1151 observations from two object communities\cite{D.Retinopathy}.
  
\textit{Connectionist Bench Data Set (Sonar):} The data set includes 208 observations of $K=2$ communities based on sonar returns collected from a metal cylinder and a cylindrically shaped rock positioned on a sandy ocean floor \cite{Sonar}.
  
\textit{QSAR Bioconcentration Classes Data Set (QSAR Bioconcentration):} The data set consists of the bioconcentration factor of 779 chemicals to determine mechanism of bioconcentration \cite{QSAR}. The data set can be partitioned into three communities.
  
\textit{Cardiotocography Data Set:} The Cardiotocography data set consists of 2126 observations of fetal cardiotocograms which can be partitioned into three communities in terms of fetal state or ten communities based on morphologic pattern \cite{Cardiotocography}.
  
\textit{Divorce Predictors Data Set:} The  data set consists of 170 observations from two object communities using divorce predictors scale \cite{DivorcePredictors}.

\subsubsection{Comparisons with Graph-based Approaches} 
The graphs for all clustering data sets are designed using Pearson's linear correlation coefficients as in Eq.~\eqref{eq:Pearson}, except for the Ionosphere and Cardiotocography data sets, where the graphs are designed as the same procedure with human gait and person identification data sets.

The estimated community numbers, for all community detection algorithms whose computation time is less than 12 hours, are summarized in Tab.~\ref{tab:tablerealclusters}. Again, the SVD is only applicable for bipartite networks. SPARCODE correctly estimates the number of clusters for all data sets and outperforms all its competitors. None of the competitor community detection algorithms is able to correctly estimate the community number of the person identification data set correctly. These overestimate it in most cases, which can be explained by the considerable number of outliers.
 \begin{figure}[!tbp]
\begin{minipage}[b]{1.0\linewidth}
  \centering
  \includegraphics[trim={5mm 5mm 5mm 0mm},clip,width=9 cm]{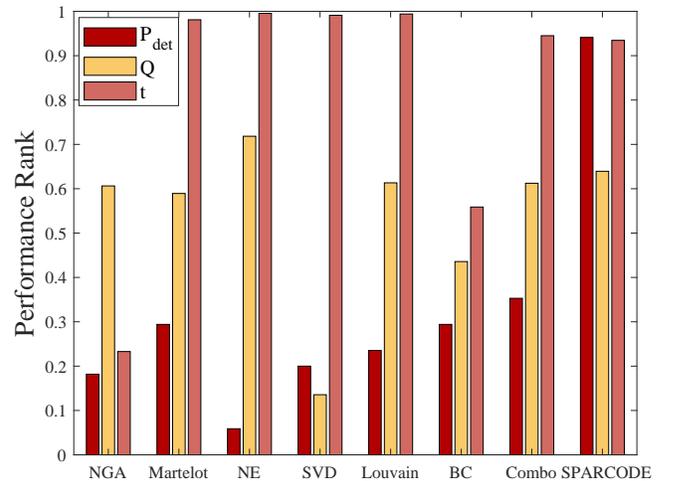}
\end{minipage}
\caption{Empirical probability of detection and average normalized performance rank of each algorithm in terms of modularity and computation time.}
\label{fig:Detailedperformancerank}
\end{figure}

\begin{table*}[h!]
\centering
\begin{tabularx}{\textwidth}{l*{8}{>{\centering\arraybackslash}X}}
\hline\hline
 & \multicolumn{7}{c}{$\hat{K}$ for Different Community Detection Methods} &  \\
\cline{2-8}
Data Set & BCE & dip-means & K. dip-means & x-means & g-means & DBSCAN & SPARCODE & K\\
\midrule
Human Gait \cite{humangait},&7 &4 & 3 & 176 &9 &6 &5 &5\\
Person Identification \cite{PersonIdentification},&4&4&3&22&10&3 &4&4\\
Iris \cite{Iris}, &3 &2& 2&31 &3 &2&3&3\\
Ionosphere \cite{Ionosphere},&2 &31 & 1 &32 &2 &2& 2&2\\
Parkinson Acoustic \cite{ParkinsonAcoustic},&1 &1 &1&47&1&1 &2&2\\
D. Retinopathy \cite{D.Retinopathy},&1&2 &2 &199 &16 &1 &2&2\\
Sonar \cite{Sonar},&1 &1 &1 &29 &1 &1&2&2\\
QSAR Bioconcentration \cite{QSAR},&5 &4&6 &174&55&1&3&3\\
Cardiotocography \cite{Cardiotocography},&8 &4 &4 &435 &51 & 2&3&3,10\\
Divorce Predictors \cite{DivorcePredictors},&1 &2 &2 &21&3 &1&2&2\\
\hline\hline
\end{tabularx}
\caption{\label{tab:table-name}Performance of cluster-based approaches on clustering data sets. The results whose computation take more than 12 hours and nontarget networks for SVD method denoted as "-".}
\label{tab:tablerealclusterbased}
\end{table*}

In addition to the accuracy in terms of the estimated number of communities, the results are also evaluated with respect to  partitioning quality and computation time. To report the modularity and computation time in the same scale as the probability of detection, the performances of all algorithms are ranked and then averaged over all real data sets. The performance measures of the SVD approach are evaluated on the bipartite graphs, only. All results are summarized in Fig.~\ref{fig:Detailedperformancerank}. As can be seen from the figure, SPARCODE achieves the best performance with 0.94 probability of detection while the nearest competitor follows with 0.35 empirical probability of detection. 

Based on quality of partition, NE achieves the best performance with averaged rank score of 0.72; SPARCODE, Louvain and Combo follow with 0.64 and 0.61 for both, respectively. Although NE shows the best performance on modularity, it has the worst performance with considerable difference in terms of its probability of detection. NE, Louvain and SVD methods are the best algorithms with respect to computation time where Martelot, Combo and SPARCODE follow with 0.98, 0.94 and 0.93, respectively. 

\begin{figure}[!tbp]
\begin{minipage}[b]{1.0\linewidth}
  \centering
  \includegraphics[trim={0mm 0mm 0mm 0mm},clip,width=9.5cm,height=7.5cm]{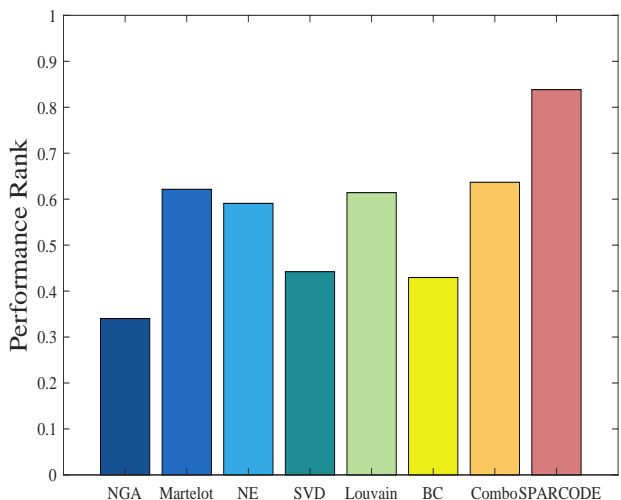}
\end{minipage}
\caption{Performance of different algorithms based on equal weights on performance metrics.}
\label{fig:overallperformancerank}
\end{figure}

To summarize the performance of graph-based community detection approaches on real data sets,
the detailed performance measures that is explained in Fig.~\ref{fig:Detailedperformancerank} are further aggregated by equally weighting each performance measure. The overall performance of eight competitor methods is summarized in Fig.~\ref{fig:overallperformancerank}. As can be seen, SPARCODE achieves the highest overall performance score of 0.84 whereas, Combo is the best competitor with a score of 0.64. Martelot, Louvain and NE follow with an overall performance score of 0.62, 0.61 and 0.59, respectively. 

\vspace{4mm}
\subsubsection{Comparisons with Cluster-based Approaches} In this section, the SPARCODE algorithm is compared with six well-known clustering methods that estimate the number of communities $K$. The community detection results of different approaches are summarized in Tab.~\ref{tab:tablerealclusterbased}. As can be seen, none of the competitor algorithms performs well in the highly contaminated human gait data set. However, generally speaking, cluster-based approaches show better performance compared to competitor graph-based approaches on person identification, iris and ionosphere data sets, except for x-means, which generally largely overestimates the number of communities.

\begin{figure}[!tbp]
\begin{minipage}[b]{1.0\linewidth}
  \centering
  \includegraphics[trim={0mm 0mm 0mm 0mm},clip,width=9.5 cm,height=7.5cm]{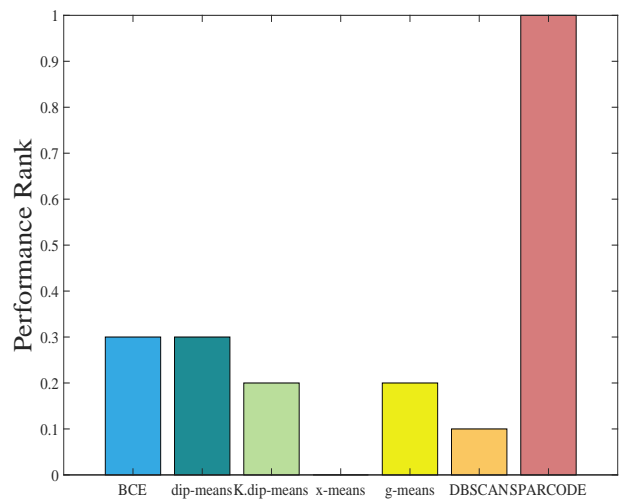}
\end{minipage}
\caption{Performance of different algorithms based on rank in terms of probability of detection.}
\label{fig:clusterbasedoverallperformancerank}
\end{figure}

The overall community detection performance of all algorithms is summarized in Fig.~\ref{fig:clusterbasedoverallperformancerank}. The SPARCODE method clearly outperforms existing cluster-based community detection approaches in terms of probability of detection and outperforms the best competitors (BCE and dip-means), which have a performance score of 0.3, by a large margin. 

\section{Conclusion}\label{sec:Conclusion}
We proposed SPARCODE, a community detection method that uses spectral partitioning based on estimating a robust and sparse graph. The level of sparseness is controlled by maximizing the modularity of the graph. SPARCODE includes a graph construction-based outlier detection method to increase robustness. Overall, when compared to both cluster-based and graph-based community detection algorithms on real and synthetic data sets, SPARCODE achieves the highest community detection performance providing a high quality of partition at a reasonable computation time compared to existing graph-based approaches.  Judging from its applicability in a large variety of problems, compatibility with different graph-based methods and success in highly contaminated data sets, SPARCODE is a promising new algorithm for performing community detection in a robust manner.

\appendices
\section{}\label{app:Scenario1}
\vspace{-1mm}
The first scenario considers a weighted graph with $n=300$ nodes that belong to seven different communities. The affinity matrix ${\qvec{W}\in\mathbb{R}^{n\times n}}$, whose entries vary between zero and one, is symmetric, zero diagonal and nonnegative. The communities are labelled as $\textnormal{C1},\dots,\textnormal{C7}$, and the parameters of similarity coefficients within the communities and between different communities are summarized in
Tab.~\ref{tab:tableScenarioo1}. The number of objects for each community is denoted as $n_{\textnormal{C}}$. The similarity coefficients are generated as $w=\mu+r\sigma$ where $r$ is a random real number distributed as $U[0,1)$.
\begin{table}[h]
\begin{center}
\begin{tabularx}{0.45\textwidth}{ssbbssssss}
\hline\hline
\cline{3-9}
& & C1 & C2 & C3 & C4 & C5 & C6 & C7 & $n_{\textnormal{C}}$\\
\midrule
\multirow{2}{4em}{C1}&$\mu$ &0.8&0.005&0.35& 0& 0& 0.25& 0& \multirow{2}{4em}{45} \\
\cline{3-9}
             &$\sigma$ &0.19&0.0045& 0.1 & - &-  &0.05  &-  & \\
\cline{1-10}                   
\multirow{2}{4em}{C2}&$\mu$ &0.005& 0.7& 0& 0.35& 0.19& 0& 0&\multirow{2}{4em}{40} \\
\cline{3-9}
                &$\sigma$ &0.0045&0.1&-&0.1& 0.15& - &-  & \\
\cline{1-10}                   
\multirow{2}{4em}{C3}&$\mu$ &0.35& 0& 0.9& 0& 0&0& 0&\multirow{2}{4em}{55}\\
\cline{3-9}
              &$\sigma$ &0.1 & -&0.09 &-& -&-& -&\\
\cline{1-10}                   
\multirow{2}{4em}{C4}&$\mu$ &0&0.35&0&0.85& 0&0&0&\multirow{2}{4em}{42} \\
\cline{3-9}
             &$\sigma$ &-&0.1&-&0.14& -&-& -& \\
\cline{1-10}                   
\multirow{2}{4em}{C5}&$\mu$ &0& 0.19&0& 0& 0.79& 0&0&\multirow{2}{4em}{37} \\
\cline{3-9}
                 &$\sigma$ &-&0.15&-&-& 0.15&-&-&\\
\cline{1-10}                   
\multirow{2}{4em}{C6}&$\mu$ &0.25&0&0&0&0& 0.85&0&\multirow{2}{4em}{46} \\
\cline{3-9}
        &$\sigma$ &0.05&-&-&-&-&0.14&-& \\
\cline{1-10}                   
\multirow{2}{4em}{C7}&$\mu$&0& 0&0& 0& 0& 0& 0.95& \multirow{2}{4em}{35} \\
\cline{3-9}
            &$\sigma$ &-&-&- &-&-&-&0.04 & \\
\hline\hline
\end{tabularx}
\caption{Similarity coefficients for seven object communities for Scenario 1.}
\label{tab:tableScenarioo1}
\end{center}
\end{table}
\vspace{-7mm}
\section{}\label{app:Scenario2}
\vspace{-1mm}
The second scenario, considers again a weighted graph with $n=300$ nodes. In contrast, the community number is $K=3$ and we investigate the effect of outliers by creating outliers which have neglibly small correlation with all communities. The communities are labelled as $\textnormal{C1},\dots,\textnormal{C3}$, and the outliers as $\textnormal{O}$, and the associated parameters of similarity coefficients within and between communities and outliers are summarized in 
Tab.~\ref{tab:tableScenarioo2}. Both the number of objects of communities and outliers are denoted as $n_{\textnormal{C}}$. The similarity coefficients are generated as $w=\mu+r\sigma$ where $r$ is a random real number distributed as $U[0,1)$ as in the first scenario.
\begin{table}[h]
\begin{center}
\begin{tabularx}{0.45\textwidth}{ssbbbbs}
\hline\hline
\cline{3-6}
& & C1 & C2 & C3 & O & $n_{\textnormal{C}}$\\
\midrule
\multirow{2}{4em}{C1}&$\mu$ &0.7&0.01&0& 0.005& \multirow{2}{4em}{66} \\
\cline{3-6}
&$\sigma$&0.29&0.29& - & 0.025 & \\
\cline{1-7}                   
\multirow{2}{4em}{C2}&$\mu$ &0.01& 0.8& 0.27& 0.005& \multirow{2}{4em}{102}\\
\cline{3-6}
                &$\sigma$ &0.29&0.19&0.1&0.025 & \\
\cline{1-7}                   
\multirow{2}{4em}{C3}&$\mu$ &0& 0.27& 0.7& 0.005& \multirow{2}{4em}{90}\\
\cline{3-6}
              &$\sigma$ &- & 0.1&0.24 &0.025&\\
\cline{1-7}                   
\multirow{2}{4em}{O}&$\mu$ &0.005&0.005&0.005&0.005&\multirow{2}{4em}{42}\\
\cline{3-6}
             &$\sigma$ &0.025&0.025&0.025&0.025  & \\
\hline\hline
\end{tabularx}
\caption{Similarity coefficients for three object communities and outliers for Scenario 2.}
\label{tab:tableScenarioo2}
\end{center}
\end{table}

\section*{Acknowledgment}
The work of A. Taştan is supported by the Republic of Turkey Ministry of National Education. The work of M. Muma has been funded by the LOEWE initiative (Hesse, 
Germany) within the emergenCITY centre and is supported by the ‘Athene Young
Investigator Programme’ of Technische Universität Darmstadt, Hesse, 
Germany.

\ifCLASSOPTIONcaptionsoff
  \newpage
\fi


\begin{thebibliography}{00}

\bibitem{l1graph} B. Cheng, J. Yang, S. Yan, Y. Fu and T. S. Huang, ``Learning with \(\ell^1\)-graph for image analysis,'' \textit{IEEE Trans. Image Process.}, vol. 19, pp. 858-866, 2009.

\bibitem{elasticnethypergraph} Q. Liu, Y. Sun, C. Wang, T. Liu and D. Tao, ``Elastic net hypergraph learning for image clustering and semi-supervised classification,'' \textit{IEEE Trans. Image Process.}, vol. 26, pp. 452-463, 2016.

\bibitem{SparseSubspace} E. Elhamifar and R. Vidal, ``Sparse subspace clustering: Algorithm, theory, and applications,'' \textit{IEEE Trans. Pattern Anal. Mach. Intell.}, vol. 35, pp. 2765-2781, 2013.

\bibitem{sparserepimage} J. Wright, A. Y. Yang, A. Ganesh, S. S. Sastry and Y. Ma, ``Robust face recognition via sparse representation,'' \textit{IEEE Trans. Pattern Anal. Mach. Intell.}, vol. 31, pp. 210-227 , 2008.

\bibitem{networkmodelling} F. Fages, S. Soliman, and N. Chabrier-Rivier, ``Modelling and querying interaction networks in the biochemical abstract machine BIOCHAM,'' \textit{J. Biol. Phys. Chem.}, vol. 4, pp. 64-73, 2004.

\bibitem{RadarConf} A. Ta{\c{s}}tan, M. Muma and A. M. Zoubir, ``An unsupervised approach for graph-based robust clustering of human gait signatures,'' in \textit{Proc. 2020 IEEE Radar Conf (accepted)}, 2020.

\bibitem{subspacelearning} G. Liu, Z. Lin, S. Yan, J. Sun, Y. Yu and Y. Ma, ``Robust recovery of subspace structures by low-rank representation,'' \textit{IEEE Trans. Pattern Anal. Mach. Intell.}, vol. 35, pp. 171-184, 2012.

\bibitem{manifold1} S. T. Roweis and L. K. Saul, ``Nonlinear dimensionality reduction by locally linear embedding,'' \textit{Sci.}, vol. 290, pp. 2323-2326, 2000.

\bibitem{manifold2} M. Belkin and P. Niyogi, ``Laplacian eigenmaps for dimensionality reduction and data representation,'' \textit{Neural Comput.}, vol. 15, pp. 1373-1396, 2003.

\bibitem{semisupervised} X. Zhu, ``Semi-supervised learning literature survey,'' \textit{Comput. Sci.}, vol. 37, pp. 63-67, 2008.

\bibitem{graphpartition1} J. Shi and J. Malik, ``Normalized cuts and image segmentation,'' \textit{IEEE Trans. Pattern Anal. Mach. Intell.}, vol. 22, pp. 888-905, 2000.

\bibitem{graphpartition2} B. Karrer and M. E. J. Newman, ``Stochastic blockmodels and community structure in networks,'' \textit{Phys. Rev. E}, vol. 83, pp. 016107, 2011.

\bibitem{NGA} M. E. J. Newman, ``Fast algorithm for detecting community structure in networks,'' \textit{Phys. Rev. E}, vol. 69, pp. 066133, 2004.

\bibitem{Martelot} E. L. Martelot and C. Hankin, ``Multi-scale community detection using stability as optimization criterion in a greedy algorithm,'' in \textit{Proc. Intl. Conf. Knowl. Discovery and Inf. Retrieval}, pp. 208-217, 2011.

\bibitem{NE} M. E. J. Newman, ``Finding community structure in networks using the eigenvectors of matrices,'' \textit{Phys. Rev. E}, vol. 74, pp. 036104, 2006.

\bibitem{SVD} S. Sarkar and A. Dong, ``Community detection in graphs using singular value decomposition,'' \textit{Phys. Rev. E}, vol. 83, pp. 046114, 2011.

\bibitem{Louvain} V. D. Blondel, J. L. Guillaume, R. Lambiotte and E. Lefebvre, ``Fast unfolding of communities in large networks,'' \textit{J. Stat. Mech: Theory Exp.}, vol. 10, pp. P10008, 2008.

\bibitem{Bayesian} J. M. Hofman and C. H. Wiggins, ``Bayesian Approach to Network Modularity,'' \textit{Phys. Rev. Lett.}, vol. 100, pp. 258701, 2008.

\bibitem{Combo} S. Sobolevsky, R. Campari, A. Belyi and C. Ratti, ``General optimization technique for high-quality community detection in complex networks,'' \textit{Phys. Rev. E}, vol. 90, pp. 012811, 2014.

\bibitem{ROBUSTNESS1} A. M. Zoubir, V. Koivunen, Y. Chakhchoukh and M. Muma, ``Robust estimation in signal processing: A tutorial-style treatment of fundamental concepts'' \textit{IEEE Signal Process. Mag.}, vol. 29, pp. 61-80, 2012.

\bibitem{zoubir-2018} A. M. Zoubir, V. Koivunen, E. Ollila and M. Muma, \textit{Robust statistics for signal processing}, Cambridge, 2018.

\bibitem{ROBUSTNESS2} B. Kad{\i}o{\u{g}}lu, {\.I}. Y{\i}ld{\i}z, P. Closas and D. Erdo{\u{g}}mu{\c{s}}, ``M-estimation-based subspace learning for brain computer interfaces'' \textit{IEEE J. Sel. Top. Signal Process.}, vol. 12, pp. 1276-1285, 2018.

\bibitem{ROBUSTNESS3} F. R. Hampel, E. M. Ronchetti, P. J. Rousseeuw and W. A. Stahel, \textit{Robust statistics: the approach based on influence functions}, John Wiley \& Sons, 2011.

\bibitem{ROBUSTNESS4} E. Ollila, and D. E. Tyler, ``Regularized $M$-estimators of scatter matrix'' \textit{IEEE Trans. Signal Process.}, vol. 62, pp. 6059-6070, 2014.

\bibitem{ROBUSTNESS5} J. Liu, and D. P.  Palomar, ``Regularized robust estimation of mean and covariance matrix for incomplete data'' \textit{IEEE Signal Process.}, vol. 165, pp. 278-291, 2019.

\bibitem{maronna-2019} R. A. Maronna, R. D. Martin, V. J. Yohai and M. Salibi{\'a}n-Barrera, \textit{Robust statistics: theory and methods (with R)}, John Wiley \& Sons, 2019.

\bibitem{ROBUSTNESS6} F. Pascal, P. Forster, J. -P. Ovarlez and P. Larzabal, ``Performance analysis of covariance matrix estimates in impulsive noise'' \textit{IEEE Trans. Signal Process.}, vol. 56, pp. 2206-2217, 2008.

\bibitem{ROBUSTNESS7} R. Couillet, F. Pascal, and J. W. Silverstein, ``The random matrix regime of {M}aronna’s {M}-estimator with elliptically distributed samples'' \textit{J. Multivariate Anal.}, vol. 139, pp. 56-78, 2015.

\bibitem{ROBUSTNESS8} V. {\"O}llerer, and C. Croux, ``Robust high-dimensional precision matrix estimation'' \textit{Modern nonparametric, robust and multivariate methods}, pp. 325-350, 2015. 

\bibitem{ROBUSTNESS9} C. A. Schroth, and M. Muma, ``Robust M-estimation based {B}ayesian cluster enumeration for real elliptically symmetric distributions,'' Online-Edition: https://arxiv.org/abs/2005.01404, 2020.

\bibitem{ROBUSTNESS10} F. K. Teklehaymanot, M. Muma and A. M. Zoubir, ``Robust {B}ayesian cluster enumeration,'' Online-Edition: https://arxiv.org/abs/1811.12337, 2018.

\bibitem{Newmanmod} A. Clauset, M. E. J. Newman and C. Moore, ``Finding community structure in very large networks,'' \textit{Phys. Rev. E}, vol. 70, pp. 066111, 2004.

\bibitem{sparsegraphest} Q. Tao, X. Huang, S. Wang, X. Xi and L. Li, ``Multiple Gaussian graphical estimation with jointly sparse penalty,'' \textit{Signal Process.}, vol. 128, pp. 88-97, 2016.

\bibitem{sparsityinclustering} H. Cheng, Z. Liu, L. Yang and X. Chen, ``Sparse representation and learning in visual recognition: Theory and applications,'' \textit{Signal Process.}, vol. 93, pp. 1408-1425, 2013.

\bibitem{LeightonRao} T. Leighton and S. Rao, ``Multicommodity max-flow min-cut theorems and their use in designing approximation algorithms,'' \textit{J. ACM}, vol. 46, pp. 787-832, 1999.

\bibitem{Geometryflows} S. Arora, S. Rao and U. Varizani, ``Geometry, flows, and graph-partitioning algorithms'' \textit{Commun. ACM}, vol. 51, pp. 96-105, 2008.

\bibitem{AppSparsestcut2} S. Arora, S. Rao and U. Varizani, ``Expander flows, geometric embeddings and graph partitioning'' \textit{J. ACM}, vol. 56, pp. 1-37, 2009.

\bibitem{Glasso} J. Friedman, T. Hastie and R. Tibshirani, ``Sparse inverse covariance estimation with the graphical {L}asso'' \textit{Biostat.}, vol. 9, pp. 432-441, 2008.

\bibitem{graphbasedfeaselect} Z. Liu, Z. Lai, W. Ou, K. Zhang and R. Zheng, ``Structured optimal graph based sparse feature extraction for semi-supervised learning'' \textit{Signal Process.}, vol. 170, pp. 107456, 2020.

\bibitem{penaltyselect1} T. Hastie, R. Tibshirani and J. Friedman, \textit{The elements of statistical learning: data mining, inference, and prediction}, Springer, 2009.

\bibitem{penaltyselect2} N. Meinshausen and P. B{\"u}hlmann, ``Stability selection,'' \textit{J. R. Stat. Soc. Ser. B 72}, vol. 72, pp. 417-473, 2010.

\bibitem{supervisedpenalty1} A. B. Owen and P. O. Perry, ``Bi-cross-validation of the SVD and the nonnegative matrix factorization,'' \textit{Ann. Appl. Stat.}, vol. 3, pp. 564-594, 2009.

\bibitem{neighborselect} N. Meinshausen and P. B{\"u}hlmann, ``High-dimensional graphs and variable selection with the lasso,'' \textit{Ann. Stat.}, vol. 34, pp. 1436-1462, 2006.

\bibitem{sparsitypenalty} K. M. Tan, D. Witten and A. Shojaie, ``The cluster graphical {L}asso for improved estimation of Gaussian graphical models,'' \textit{Comput. Stat. Data Anal.}, vol. 85, pp. 23-36, 2015.

\bibitem{NetworksNewman} M. E. J. Newman, \textit{Networks}, Oxford, 2010.

\bibitem{SparseEnough} D. L. Donoho, ``For most large underdetermined systems of linear equations the minimal $\ell^{1}$-norm solution is also the sparsest solution,'' \textit{Commun. Pure Appl. Math.}, vol. 59, pp. 797-829, 2006.

\bibitem{glassolimitation} R. Mazumder and T. Hastie, ``The graphical {L}asso: New insights and alternatives'' \textit{Electron. J. Stat.}, vol. 6, pp. 2125, 2012.

\bibitem{AppSparsestcut1} S. Arora, E. Hazan and S. Kale, ``$O\sqrt{\textnormal{log}n}$ approximation to sparsest cut in $\Tilde{O}(n^2)$ time,'' \textit{SIAM J. Comput.}, vol. 39, pp. 1748-1771, 2010.

\bibitem{Laplacianeigenmaps} M. Belkin and P. Niyogi, ``Laplacian eigenmaps and spectral techniques for embedding and clustering'' \textit{Adv. Neural Inf. Process. Syst.}, pp. 585-591, 2002.

\bibitem{Fiedler} M. Fiedler, ``A property of eigenvectors of nonnegative symmetric matrices and its application to graph theory,'' \textit{Czechoslovak Math. J.}, vol. 25, pp. 619-633, 1975.

\bibitem{Fiedlercut} D. A. Spielman and S. -H. Teng, ``Spectral partitioning works: Planar graphs and finite element meshes,'' in \textit{Linear Algebra Appl.}, vol. 421, pp. 284-305, 2007.

\bibitem{Newman2002} M. Girvan and M. E. J. Newman, ``Community structure in social and biological networks,'' in \textit{Proc. Natl.
Acad. Sci.}, vol. 99, pp. 7821-7826, 2002.

\bibitem{Newman2006}  M. E. J. Newman, ``Modularity and community structure in networks,'' in \textit{Proc. Natl.
Acad. Sci.}, vol. 103, pp. 8577-8582, 2006.

\bibitem{BayesianTekle} F. K. Teklehaymanot, M. Muma and A. M. Zoubir, ``Bayesian cluster enumeration criterion for unsupervised learning,'' \textit{IEEE Trans. Signal Process.}, vol. 66, pp. 5392-5406, 2018.

\bibitem{dipmeans} A. Kalogeratos and A. Likas, ``Dip-means: An incremental clustering method for estimating the number of clusters,'' in \textit{Proc. Adv. Neural Inf. Process. Syst.}, pp. 2393-2401 , 2012.

\bibitem{xmeans} D. Pelleg and A. Moore, ``X-means: Extending K-means with efficient estimation of the number of clusters,'' in \textit{Proc. 17th Int. Conf. Mach. Learn. (ICML)}, pp. 727-734 , 2000.

\bibitem{gmeans} G. Hamerly and E. Charles, ``Learning the K in K-Means,'' in \textit{Proc. 16th Int. Conf. Neural Inf. Process. Syst. (NIPS)}, pp. 281-288 , 2004.

\bibitem{DBSCAN} M. Ester, H. -P. Kriegel, J. Sander and X. Xu, ``A density-based algorithm for discovering clusters in large spatial databases with noise,'' in \textit{Proc. 2nd Int. Conf. Knowl. Discovery Databases Data Min.}, pp. 226-231, 1996.

\bibitem{Karate} W. W. Zachary, ``An information flow model for conflict and fission in small groups,'' \textit{J. Anthropological Res.}, vol. 33, pp. 452-473, 1977.

\bibitem{Dolphins} D. Lusseau, K. Schneider, O. J. Boisseau, P. Haase, E. Slooten, and S. M. Dawson, ``The bottlenose dolphin community of doubtful sound features a large proportion of long-lasting associations,'' \textit{Behav. Ecol. Sociobiol.}, vol. 54, pp. 396-405, 2003.

\bibitem{PoliticalBlogs} L. A. Adamic and N. Glance, ``The political blogosphere and the 2004 US election: divided they blog,'' in \textit{Proc. 3rd Intl. Workshop Link Discovery Res.}, pp. 36-43, 2005.

\bibitem{Jazz} P. M. Gleiser and L. Danon, ``Community structure in Jazz,'' \textit{Adv. Complex Syst.}, vol. 6, pp. 565-573, 2005.

\bibitem{Carpinteria} K. D. Lafferty, A. P. Dobson and A. M. Kuris, ``Parasites dominate food web links,'' in \textit{Proc. Natl. Acad. Sci.}, vol. 103, pp. 11211-11216, 2006.

\bibitem{Celegans} B. L. Chen, D. H. Hall and D. B. Chklovskii, ``Wiring optimization can relate neuronal structure and function,'' in \textit{Proc. Natl. Acad. Sci.}, vol. 103, pp. 4723-4728, 2006.

\bibitem{humangait} A. -K. Seifert, M. Amin and A. M. Zoubir, ``Toward unobtrusive in-home gait analysis based on radar micro-{D}oppler signatures,'' \textit{IEEE Trans. Biomed. Eng.}, vol. 66, pp. 1-11, 2019.

\bibitem{radarinfo} Ancortek  Inc., ``SDR-KIT  2400AD,''  [Online].  Available:  http://ancortek.com/sdr-kit-2400ad. Accessed on: Jul. 13, 2017.

\bibitem{PersonIdentification} F. K. Teklehaymanot, A. -K. Seifert, M. Muma, M. G. Amin and A. M. Zoubir, ``Bayesian target enumeration and labeling using radar data of human gait,'' in \textit{Proc. 26th European Signal Process. Conf. (EUSIPCO)}, pp. 1342-1346, 2018.

\bibitem{Iris} R. A. Fisher, ``The use of multiple measurements in taxonomic problems,'' \textit{Ann. Eugenics}, vol. 7, pp. 179–188, 1936.

\bibitem{Ionosphere} V. G. Sigilitto, S. P. Wing, L. V. Hutton and K. B. Baker, ``Classification of radar returns from the ionosphere using neural networks,'' \textit{Johns Hopkins APL Tech. Dig.}, vol. 10, pp. 262–266, 1989.

\bibitem{ParkinsonAcoustic} L. Naranjo, C. J. Perez, Y. Campos-Roca and J. Martin, ``Addressing voice recording replications for Parkinson’s disease detection,'' \textit{Expert Syst. Appl.}, vol. 46, pp. 286-292, 2016.

\bibitem{D.Retinopathy} B. Antal and A. Hajdu, ``An ensemble-based system for automatic screening of diabetic retinopathy,'' \textit{Knowledge Based Syst.}, vol. 60, pp. 20-27, 2014.

\bibitem{Sonar} R. P. Gorman and T. J. Sejnowski, ``Analysis of hidden units in a layered network trained to classify sonar targets,'' \textit{Neural Networks}, vol. 1, pp. 75-89, 1988.

\bibitem{QSAR} F. Grisoni, V. Consonni, M. Vighi, S. Villa and R. Todeschini, ``Investigating the mechanisms of bioconcentration through QSAR classification trees,'' \textit{Environ. Int.}, vol. 88, pp. 198-205, 2016.

\bibitem{Cardiotocography} D.Ayres-de-Campos, J. Bernardes, A. Garrido, J. Marques-de-Sa and L. Pereira-Leite, ``SisPorto 2.0: a program for automated analysis of cardiotocograms,'' \textit{J. Maternal-Fetal Med.}, vol. 9, pp. 311-318, 2000. 

\bibitem{DivorcePredictors} M. K. Y\"{o}ntem, K. Adem, T. \.{I}lhan and S. K{\i}l{\i}{\c{c}}arslan, ``Divorce prediction using correlation based feature selection and artificial neural networks,'' \textit{J. Nev{\c{s}}ehir Hac{\i} Bekta{\c{s}} Veli University SBE}, vol. 9, pp. 259-273, 2019. 


\end{thebibliography}
\end{document}